\documentstyle[12pt,aasms4]{article}
\everymath{\rm}
\everydisplay{\rm}
\received{}
\revised{}
\accepted{}
\cpright{}{}\journalid{}{}
\articleid{}{}
\paperid{}
\slugcomment{Accepted for publication in the Astrophysical Journal}
\lefthead{Henry, Kwitter, \& Bates}
\righthead{Carbon In Planetary Nebulae}
\begin{document}
\title{A New Look At Carbon Abundances In Planetary Nebulae. IV. Implications For Stellar Nucleosynthesis}
\author{R.B.C. Henry\footnote{Visiting Astronomer, Kitt Peak
National Observatory, National Optical Astronomy Observatories,
which is operated by the Association of Universities for Research in
Astronomy, Inc. (AURA) under cooperative agreement with the
National Science Foundation.}}
\affil{Department of Physics \& Astronomy, University of
Oklahoma, Norman, OK  73019; henry@mail.nhn.ou.edu}
\author{K.B. Kwitter$^1$ \& J.A. Bates}
\affil{Department of Astronomy, Williams College, Williamstown,
MA  01267; kkwitter@williams.edu}

\begin{abstract}

This paper is the fourth and final report on a project designed to
study carbon abundances in a sample of planetary nebulae representing a
broad range in progenitor mass and metallicity.  We present newly
acquired optical spectrophotometric data for three Galactic planetary
nebulae IC~418, NGC~2392, and NGC~3242 and combine them with UV data
from the IUE Final Archive for identical positions in each nebula to
determine accurate abundances of He, C, N, O, and Ne at one or more
locations in each object. We then collect abundances of these
elements for the entire sample and compare them with theoretical
predictions of planetary nebula abundances from a grid of intermediate
mass star models. We find some consistency between observations and
theory, lending modest support to our current understanding of
nucleosynthesis in stars below 8~M$_{\odot}$ in birth mass. Overall, we believe that observed abundances agree with theoretical predictions to well within an order of magnitude but probably not better than within a factor of 2 or 3. But even this level of consistency between observation and theory enhances the validity of published intermediate-mass stellar yields of carbon and nitrogen in the study of the abundance evolution of these elements.

\end{abstract}


\section{Introduction}

Identifying the origin of most heavy elements, i.e. those elements heavier than helium, has long been one of the goals of research into the chemical evolution of galaxies. While production rates of most isotopes are not known to better than a factor of two (Woosley \& Weaver 1995; Nomoto et al. 1997), there is nevertheless little argument over the idea that massive stars (M$>$8~M$_{\odot}$) are the principal, indeed in most cases the sole, source of these elements, since it is only within these stars that temperatures reach the levels necessary for forging them. 

In the case of carbon and nitrogen, however, the origins are more ambiguous. Intermediate mass stars (IMS; 1$\le$M$\le$8~M$_{\sun}$) possess temperatures in their cores and shell-fusing regions which permit carbon production through helium burning as well as nitrogen production via the CNO cycle, thus making these stars potentially responsible for some portion of the production of these two elements. Recent IMS models and yield calculations by van~den~Hoek \& Groenewegen (1997) and by Marigo, Bressan, \& Chiosi (1996; 1998) have provided modern updates to the original work of Renzini \& Voli (1982) by using current opacities and more sophisticated mass loss relations. The collective results of these theoretical studies strongly suggest that significant amounts of both carbon and nitrogen are produced in IMS. This theoretical finding is supported by observed supersolar levels of carbon and nitrogen in planetary nebulae (PNe), which are objects formed during the asymptotic giant branch stage of IMS evoltuion and which comprise gas containing stellar nucleosynthetic products which have been mixed to the stellar surface and expelled into the interstellar medium (see {\S}4.2 for a brief discussion on IMS evolution).

At the same time, however, models of massive stars suggest that they too synthesize and expel significant amounts of carbon and nitrogen.
Massive star yields have been calculated by Woosley \& Weaver (1995), Nomoto et al. (1997), and Maeder (1992), where the last author included metallicity-sensitive mass loss in his models. When the relative number of stars in each mass range is accounted for, the general impression from yield predictions over the whole stellar mass range is that nitrogen originates predominantly in IMS but that carbon may come both from IMS and massive stars in roughly equal amounts. If Maeder's yields are used, carbon is mostly produced in massive stars, as demonstrated by the chemical evolution models of Prantzos, Vangioni-Flam, \& Chauveau (1994) and recently supported by observations by Gustafsson et al. (1999).

We began a project in 1993 whose purpose was to determine accurate carbon abundances for a sample of roughly 20 planetary nebulae (PNe) representing a broad range in progenitor metallicity and mass in order to confront published stellar models and their predictions of PN abundances. Since these same calculations also provide total chemical yields (not just what is expelled during the PN phase), this comparison would allow us to assess the validity of these yields. We planned our project in order to exploit the newly-available Final Archive IUE database, containing the sum of UV observations gathered over the satellite's life, and which had been re-reduced in a consistent manner. Our plan was to combine the UV data with optical data for the same nebular positions to be able to measure numerous spectral line strengths and obtain abundances for carbon as well as for helium, nitrogen, oxygen, and neon.  Once abundances were established our intension was to compare these values with PN abundance predictions taken from models of IMS.

This paper is the fourth and final installment in our series which focuses on carbon in PNe. Each previous paper, Henry, Kwitter, \& Howard (1996; Paper~I) and Kwitter \& Henry (1996, Paper~II; 1998, Paper~III), along with a dedicated paper on NGC~7293, Henry, Kwitter, \& Dufour (1999) has provided the abundance results for a specific subset of objects in our sample and thus served as a progress report as we worked through our list of objects. In this paper, we report on the analysis of our three remaining objects, IC~418, NGC~2392, and NGC~3242, present a summary of the abundances for the entire sample, and finally compare our sample results with the predicted PN abundances from the models of van~den~Hoek \& Groenewegen (1997) and Marigo, Bressan, \& Chiosi (1996).

Section 2 discusses the optical and UV observations and the reduction techniques employed, {\S}3 provides the results of the abundance calculations, {\S}4 is a discussion of the results for the entire sample, including a comparison with theoretical calculations, and a project summary is given in {\S}5.

\section{Observations and Reductions}

\subsection{Optical Observations}

Our study requires optical data along the same line-of-sight as that of the available UV data in the IUE archives. Since such data were not available for our current objects, IC~418, NGC~2392, and NGC~3242, new optical observations
were obtained at KPNO during 6-8 December 1996 with the
Goldcam CCD spectrometer attached to the 2.1m telescope. The chip was a Ford 3K $\times$ 1K
CCD with 15$\mu$ pixels. We used a slit that was 5$\arcsec$ wide and
extended 285$\arcsec$ in the E-W direction, with a spatial scale of
0$\farcs$78/pixel. With a combination of two gratings, we obtained
spectral coverage from 3700-9600\AA\ with overlapping coverage from
$\sim$5750 - 6750\AA.  Wavelength dispersion was 1.5 \AA/pixel
($\sim$8 \AA\ FWHM resolution) for the blue, 1.9 \AA/pixel ($\sim$10
\AA\ FWHM resolution) for the red.  Table~1A lists the slit position offsets in arcseconds with respect to the central star, along with the the exposure times in seconds for the blue and red grating configurations. Note that four positions were observed in NGC~2392 to match the four IUE lines-of-sight.
We obtained the usual bias and twilight
flat-field frames each night, along with HeNeAr comparison spectra for
wavelength calibration and standard star spectra for sensitivity
calibration.  The thinned red chip produces interference fringes
visible in the red.  In our red spectra the fringes appear at the
$\pm$1\% level at $\sim$7500\AA\ and increase in amplitude with
increasing wavelength: $\pm$1.5\% at 8000\AA, $\pm$4.5\% at 8500\AA,
$\pm$6\% at 9000\AA. Even at their worst, {\it i.e.\/}, at
$\sim$$\lambda$9500, the longest wavelength we measure, the fringe
amplitude reaches only about $\pm$7\%. Internal quartz flats were
taken at the position of each object both before and after the object
integrations in anticipation of removing the fringes during data
reduction. More noise was introduced in this process than was removed;
we therefore decided to leave the fringes untouched, and to accept
this additional uncertainty in our line intensities longward of
$\sim$7500\AA.

The original spectra were reduced in the standard fashion using
IRAF\footnote{IRAF is distributed by the National Optical Astronomy
Observatories, which is operated by the Association of Universities
for Research in Astronomy, Inc. (AURA) under cooperative agreement
with the National Science Foundation.}. Employing tasks in the {\it
kpnoslit\/} package, these two-dimensional spectra were converted to
one dimension by extracting a specific section along the slit. The
location of the extracted section was chosen to maximize the overlap
with the IUE aperture.

\subsection{UV Data}

All UV spectra used for this project were obtained from the IUE Final
Archive.  Spectra in the Final Archive were systematically and
uniformly re-processed by IUE staff using the NEWSIPS algorithms, and
they represent the best available calibration of these data.  We have
used all short-wavelength (SWP), low-dispersion, and large-aperture
(21$\farcs$7$\times$9$\farcs$1) spectra. Table~1B lists the SWP number, slit position offsets in arcseconds with respect to the central star, slit position angle in degrees, and exposure time in seconds for the three PNe reported on
here.

\subsection{Slit Positions}

During our optical observations, we placed the Goldcam slit in each target PN as close as possible to
the location of the best IUE observations for which detailed
positional information was available. Since the position angle of the
Goldcam slit is fixed at 90$\arcdeg$ while the IUE aperture position angle
is not, the quality of the overlap varies and will be described below
for each object. We also note that because of the 2:1 relative slit
widths, the largest possible overlap of the Goldcam slit onto the IUE
slit is $\sim$50\%.

For each object we now describe the IUE and optical observations with
regard to slit position. In general, our optical slit N-S offsets from the
central star or the center of the nebula match the IUE N-S offsets;
where they differ it is because we chose to avoid a bright star that
would have fallen on the slit. The E-W component (if any) of the IUE
offset is matched in the extraction process that creates a
one-dimensional spectrum from the appropriate portion of the
two-dimensional spectrum.

{\it IC~418}: The IUE spectra were centered 2$\arcsec$ N and
7$\arcsec$ E of the central star; the Goldcam slit was positioned
5$\arcsec$ N. The position angles of both IUE spectra are
341$\arcdeg$, leading to a fair overlap with our optical slit.

{\it NGC~2392}: The position angle for all of the IUE spectra of
NGC~2392 was 171$\arcdeg$, providing fair overlap with the Goldcam
slit.  The IUE aperture for position A was centered 15$\arcsec$ S of
the central star; the corresponding Goldcam position was
14$\arcsec$ S. The position B IUE aperture offset was 15$\arcsec$ N; the
Goldcam slit was placed 14$\arcsec$ N. IUE spectra for positions C and
D were 8$\arcsec$ E and 8$\arcsec$ W of the central star,
respectively. Optical spectra for these positions were extracted from
E and W portions of the same two-dimensional spectrum centered on the
central star.

{\it NGC~3242}: Both the IUE aperture and the Goldcam slit were offset
8$\arcsec$S of the central star. The position angle of the IUE aperture
was 257$\arcdeg$, almost parallel to the Goldcam slit; therfore the
overlap was the best possible.

\subsection{Line Strengths}

Strengths of all optical and UV lines were measured using {\it splot}
in IRAF and are reported in Table~2A.  Fluxes uncorrected for reddening
are presented in columns labelled F($\lambda$), where these flux values
have been normalized to H$\beta$=100 using our observed value of
F$_{H\beta}$ shown in the third row from the bottom of the table.
These line strengths in turn were corrected for reddening by assuming
that the relative strength of H$\alpha$/H$\beta$=2.86 and computing the
logarithmic extinction quantity $c$ shown in the penultimate line of
the table.  Values for the reddening coefficients, f($\lambda$), are
listed in column~(2), where we employed Seaton's (1979) extinction
curve for the UV and that of Savage \& Mathis (1979) for the optical.

Because of the imperfect spatial overlap between the optical and IUE
observations, a final adjustment was made by multiplying the IUE line
strengths by a merging factor that was determined from either the
theoretical ratio of the He~II lines $\lambda$1640/$\lambda$4686 or the
carbon lines C~III]~$\lambda$1909/C~II~$\lambda$4267.  The calculation
of the merging factors was described in the Appendix of Paper III;
values are listed in the last row of Table~2A.

The columns headed I($\lambda$) list our final, corrected line
strengths, again normalized to H$\beta$=100.  In general, intensities
of strong lines have uncertainties $\le$10\%; single colons indicate uncertainties up
to $\sim$25\%, and double colons denote doubtful detections with
uncertainties $\ge$50\%.

As a check on the accuracy of our final line strengths, we compare
observed and theoretical values for a number of line ratios which are
set by atomic constants in Table~2B. The first column lists the ratio,
the second column the theoretical value, and the following six columns
give the observed ratios for each location observed. Agreement is
reasonable for all but the ratio for [Ne~III], whose value may be
affected by improper subtraction of H$\epsilon$. The closeness of the
other ratios to their theoretical value seems to confirm our general
claim that strong lines have an uncertainty of $\pm$10\%, while weaker lines are a bit less certain.

\section{Results}

\subsection{Abundance Calculations}

We determined abundances of He, O, C, N, and Ne for each observed line-of-sight listed in Tables~1A,B.
The heart of our method for determining abundances is the standard one in which abundances of observable ions for an element are first determined using a 5-level atom calculation for each ion. Then these ionic abundances are summed together and multiplied by an ionization correction factor (ICF) which adjusts the sum upward to account for unobservable ions. Finally, this product is in turn multiplied by a model-determined factor $\xi$ which acts as a final correction to our elemental abundance. The method may be represented mathematically as follows:
\begin{equation} 
{{N_X}\over{N_{H^+}}} = \left\{{\sum^{obs}}{{N_i}\over{N_{H^+}}}\right\} \cdot ICF(X) \cdot \xi(X),  
\end{equation}
where
\begin{equation}
\xi(X) = {{input~model~abundances}\over{output~model~abundances}}.
\end{equation}
In eq.~1, $N_X$, $N_i$, and $N_{H^+}$ are the number abundances of element $X$, the observable ions of that element, and of $H^+$, respectively. Formulas for the ICFs come from Kingsburgh \& Barlow (1994). Each fraction in the summation is determined from the the relevant emission line strength, $I_{\lambda}$, and a volume emissivity. 
The $\xi$ factor is a modification which we introduced in Paper~I. To determine its value, we calculate a detailed photoionization model for each line-of-sight we observe in an object, matching an observed set of diagnostic line ratios which are sensitive to physical properties of the nebula such as electron temperature and density, and certain important elemental abundances. Standard model input includes central star properties along with a set of {\it input model abundances}. The model output line strengths are then used as input into our 5-level atom program to infer a set of {\it output model abundances}. Then, by assuming that $\xi$ is
identical for both the model and the actual nebula, we perform a final correction which is tailored for the specific PN being studied.  This method has been discussed in full detail most recently in Henry, Kwitter, \& Dufour (1999), and the reader is referred to that paper for more information.

The results of our abundance determinations are reported in Tables~3-5. Table~3 contains the ionic abundances and ICFs which we calculated from the line strengths in Table~2A for each line-of-sight position. Tables~4A and 4B are related to the calculation of $\xi$. Table~4A shows, in the upper panel, the ten diagnostic ratios in the first column, followed alternately by the values observed at each line-of-sight and the values calculated with a photoionization model. Input parameters for each model are shown in the lower panel. The first two of these are the central star's effective temperature and luminosity, followed by the average nebular electron density, the star-inner cloud distance, and the distance between the star and the outer edge of the nebula. The final seven rows give input abundance information for each model. We emphasize that the input values for the models, especially the elemental abundances, do not necessarily correspond to the actual nebular quantities. Rather they are the result of continuous adjustments until a match of diagnostic ratios is obtained. Our goal in calculating the models was to produce ratio values within roughly 0.15dex of the observed ones. The only major discrepancy in this regard is the log~I$_{HeII}$/I$_{HeI}$ ratio for NGC~3242. No amount of tinkering with the input parameters was able to improve the situation. We point out that the ratio for [O~II]/[O~III] above it, like the helium ratio, is sensitive to nebular excitation, and its observed and calculated values agree nicely.

Table~4B lists the values of $\xi$ computed from the models. Notice that generally $\xi$ is within 20\% of unity, which would be the value of $\xi$ if abundances determined from model output agreed exactly with the input abundances.

\subsection{Derived Abundances, Temperatures, and Densities}

Our final abundances are listed by object (or position within the object in the case of NGC~2392) in Table~5\footnote{We plan to treat sulfur and argon abundances in a separate paper, even though line strengths for these elements are available for analysis. Thus we have not calculated abundances for these elements here.}. Included in Table~5 are up to five different electron temperatures along with [S~II] densities for each object. Note that we have listed separately the results for the four positions in NGC~2392 along with arithmetic averages for that object.{\footnote{We emphasize that all of our abundances, electron temperatures, and densities are determined using the 5-level atom calculation referred to above (and described in detail in Henry et al. 1999) with the abundances determined using the scheme summarized by eqs.~1 and 2.}} The last column contains solar abundances from Grevesse, Noels, \& Sauval (1996). We estimate that our abundance ratios are uncertain by plus or minus the values in parentheses. In Figure~1 we show the abundance results graphically for each object (four points for NGC~2392), where the vertical value is either the algebraic difference between the object and the sun in the case of helium or log(X)-log(X$_{\sun}$) for the other four ratios. Our data are shown with filled symbols, while comparison abundances are shown with open symbols. Sources for the latter are given in the figure caption. Our abundance determinations are in reasonable agreement with previously published ones. Note, however, that Ne/O may be below solar in IC~418, which is interesting because this ratio is known to be quite constant in most PNe and H~II regions (Henry \& Worthey 1999).

We present five different electron temperatures for the objects in which the required line strengths were available. Uncertainties for [O~III] and [N~II] temperatures are $\pm$200K, $\pm$500K for [O~II], [S~II], and [S~III] temperatures, and $\pm$200cm$^{-3}$ for electron densities. For any one object the temperatures are roughly consistent with the exception of the [O~II] temperature, which is unrealistically high for IC~418 and rather low for the other two objects. It appears generally that the singly ionized zones, i.e. those of N$^+$, O$^+$, and S$^+$ are cooler than the zones containing O$^{+2}$ and S$^{+2}$, as is often the case due to the dilution of central star ionizing radiation at greater distances. Our temperatures and densities agree well with previously published values by Hyung et al. (1994) for IC~418, Barker (1991) for NGC~2392, and Barker (1985) for NGC~3242. We point out that in the abundance calculations above we employed the [O~III] temperatures consistently for the abundances of the doubly ionized species, while we used [N~II] temperatures for the singly ionized species.

We now place the results presented in this section in the context of our entire sample in an attempt to reach general conclusions and identify trends about element synthesis in PN progenitor stars.

\section{Discussion: Consideration Of The Entire Sample}

\subsection{The Reliability Of Our Abundances}

The first column of Table 6 provides a list of the 20 PNe which we
have now analyzed; two objects on the original list of 22, PN~06-41.1
and YM~29, were ultimately dropped due to lack of good optical or UV data.
The second column lists the Peimbert classification type as defined in
Peimbert (1978). Columns 3 through 7 give the abundance ratios as
determined in Papers I, II, III, the paper dealing exclusively with
NGC~7293 (Henry, Kwitter, \& Dufour 1999), and the current paper. For
comparative purposes, abundance ratios for the sun (Grevesse, Noels, \& Sauval 1996) and the Orion Nebula
(Esteban et al. 1998) are presented. Each abundance value is followed
in parentheses by a number representing a plus or minus uncertainty
estimate. Columns 8 and 9 give derived [O~III] and [N~II] electron
temperatures in units of 10$^3$K, while Column~10 provides the derived
[S~II] electron densities in units of cm$^{-3}$. The final column
indicates the paper in our series in which the spectral data are presented and abundances, temperatures, and densities derived.

We believe that the abundances we have determined for the objects in our sample represent the most accurate currently available for these planetary nebulae.  We make this claim based upon the following arguments.

\begin{itemize}

\item {\it Use of Final Archive IUE data}. The IUE Final Archive database contains data which have been reduced using state-of-the-art techniques developed during the life of the IUE satellite. All data were reduced with the same set of algorithms. We then remeasured all of the relevant fluxes ourselves.

\item {\it Improved slit alignment}. As already stated, consistency between IUE and optical slit positions was considered critical, and in many cases when optical data were unavailable we obtained our own to ensure good alignment.

\item {\it Merging UV and optical data}. For nearly all spatially-resolved objects in our sample we were able to merge their UV and optical spectra by forcing the observed ratio of He~II $\lambda$1640/$\lambda$4686 to be equal to its theoretical value. A secondary ratio of C~III] $\lambda$1909/C~II $\lambda$4267 was found to give reasonably good results when one or both of the helium lines was unavailable.

\item {\it Consistent abundance routine}. Abundances were calculated for our entire sample using the same procedures and reddening law, and except for a few cases, the same atomic database. This optimizes homogeneity in the results.

\item {\it Modelling of line-of-sight properties}. For each object a photoionization model was calculated to match a predetermined set of line strength ratios chosen for their sensitivity to nebular properties such as abundances, temperatures, and densities. The models were used to determine a correction value that presumably adjusts for effects of processes such as charge exchange and dielectronic recombination which are not easily accounted for by ionization correction factors.

\end{itemize}
{\it Caveat:} The accuracy of abundances in emission-line systems, and
particularly in planetary nebulae, is threatened by the proposed
existence of small scale temperature fluctuations along the
line-of-sight, first described by Peimbert (1967).  In this picture, an
electron temperature measured with forbidden lines is actually
overestimated when fluctuations are present but ignored.  This in turn
causes an underestimation of an abundance ratio when it's based upon a
forbidden/permitted line ratio.  The effect upon ratios determined
completely by forbidden or by recombination lines is minimal, since
temperature effects cancel in the first case and are small in the
second.  Oft-cited evidence of temperature fluctuations is the
finding that nebular oxygen abundances in the solar neighborhood are
roughly a factor of two below solar; indeed our O/H values reported
below are systematically below solar. But our results are supported by
B star oxygen abundances made by Smartt \& Rolleston (1997) close to
the sun and are also consistent with nebular levels.  Thus, the
evidence is ambiguous.  Temperature fluctuations have been used to
explain, among many other things, the significant discrepancy in
planetary nebula carbon abundances (Peimbert, Torres-Peimbert, \&
Luridiana 1995), where those determined using C~II
$\lambda$4267/H$\beta$, say, are often several times greater than
abundances inferred from C~III] $\lambda$1909/H$\beta$. Esteban et al.
(1998) found the effect of temperature fluctuations to be small in the
Orion Nebula, while Liu (1998) found a large effect in the planetary
nebula NGC~4361, although in their case it was still insufficient for
explaining the discrepancy between carbon abundances from recombination
and collisionally excited lines.  The issue of temperature fluctuations
is an important one, albeit unresolved. Further details can be found in
Peimbert (1995), Liu et al.  (1995), Mathis, Torres-Peimbert, \&
Peimbert (1998), Stasi{\'n}ska (1998), and Liu et al. (1999). To
summarize, in our present studies, O/H should be affected the most,
while He/H, C/O, N/O, and Ne/O should be unaffected, since temperature
effects are small or cancel in the calculation of these ratios. Thus,
if temperature fluctuations are indeed relevant, then our O/H values
represent lower limits.

We first compare our abundances with previously published results from other investigators for the same objects. Fig.~2 shows plots in separate panels of comparison (vertical axis) versus our abundances (horizontal axis) for the five measured ratios He/H, O/H, C/O, N/O, and Ne/O, where the last four are expressed logarithmically. The sources for the comparison abundances for each object are given in the figure caption. A diagonal line in each panel shows the locus of one-to-one correspondence, and error bars show typical uncertainties. 

Agreement is reasonable for all five ratios considering the uncertainties. Note that all of our C/O ratios were computed from collisionally excited lines as were all but four comparison ratios. Those four comparisons for which recombination lines were used are indicated by an x within the circle symbol. We briefly mention the three extreme objects indicated in the O/H, C/O and N/O panels. In the O/H panel, the one apparent outlier is BB1 (identified in Fig.~2), where our abundance is roughly twice that found by Pe{\~n}a et al. (1991). This difference can be traced to a larger measured value for [O~III] $\lambda$4363/$\lambda$5007 by Pe{\~n}a et al., translating to a higher electron temperature and a lower oxygen abundance. In the C/O panel, the most extreme outlier is IC~4593, for which Bohigas \& Olgu{\'i}n (1996) used recombination lines to derive a C/O value. Often, C/O is found to be significantly larger when recombination lines are used compared with results using collisionally excited lines. In fact our own C/O determination for IC~4593 using our measurements of C~II $\lambda$4267 produces a value three times greater than our collisional value (see Paper~III), consistent with the Bohigas \& Olgu{\'i}n result. Finally, we find a somewhat greater value for N/O for IC~3568 than does Perinotto (1991). Our value matches his prior to the application of our $\xi$ correction and so it is that correction which is causing the discrepancy. Harrington \& Feibelman (1983) employed a model analysis to obtain an N/O ratio which agrees well with our value prior to the application of the correction factor.

Finally, for several of our objects we presented carbon abundances determined from the strength of the C~II recombination line at 4267~{\AA}. We often found that this approach implied a carbon abundance several times greater than that obtained using the collisional lines, consistent with the findings of other investigators (see the {\it caveat} above). This discrepancy has been discussed extensively in the literature (Rola \& Stasi{\'n}ska 1994; Peimbert, Torres-Peimbert, \& Luridiana 1995), but the problem is currently unresolved.

We now proceed to the culmination of the project, which is to use our abundances to test, to the extent that the size of our sample will allow, the applicability of theoretical stellar yields by comparing the predicted PN abundances which these same theoretical calculations predict with our observed values. 

\subsection{IMS Nucleosynthesis: Models Versus Observations}

Intermediate mass stars range in mass from 1-8M$_{\odot}$,
representing roughly 45\% of all stars that form, according to a
Salpeter (1955) initial mass function.  During their lifetimes IMS may
experience up to three dredge-up phases during which fusion products
are mixed into the outer portion of the star and alter the original
composition there in ways that are dependent upon stellar mass and
metallicity.  As described by Iben (1995), the first phase occurs
during the red giant stage when CN cycle products are mixed out to the
stellar surface, causing the levels of N to rise and of C to fall.  In
the second phase, predicted to occur early in the AGB stage and only in
stars in the 2.5-8M$_{\sun}$ range, CNO products are dredged up into
the outer stellar regions, raising He and N and lowering C and O.
Finally, on the AGB the evolved star experiences thermal pulsations,
and the products of He burning, such as C, are dredged up.  Limited CN
burning may also convert some C to N. Current understanding of IMS evolution indicates that while these
stars are on the AGB, they eject a large portion of their outer
envelope, including some of the He, C and N synthesized within the
star, as they form planetary nebulae.  A recent summary of AGB
evolution and element production appears in Lattanzio (1999).

Compilations of observed abundances in PNe, such as those by Henry (1990) and Perinotto (1991) provide strong evidence that IMS synthesize
He, C, and N. We can infer that directly by comparing abundance patterns in our PN sample with patterns in the interstellar medium, i.e. H~II regions and stars.
Figures~3A,B,C show log(C/O), log(N/O), and log(Ne/O) versus 12+log(O/H) for our PN sample (filled circles) along with Galactic and extragalactic H~II region (open circles) data compiled and described in Henry \& Worthey (1999) and F and G star (open squares; Fig.~3A only) data from Gustafsson et al. (1999). Also shown in A. and B. are the positions for the sun (S; Grevesse et al. 1996), Orion (O; Esteban et al. 1998), and M8 (M; Peimbert et al. 1993; A. only). Representative uncertainties for all of the data are shown with a set of error bars.

In Fig.~3A, note that in contrast to the relatively close correlation between C and O displayed by
the H~II regions and stars, there is no such relation indicated for PNe.
In fact the range in carbon is over 2.5 orders of magnitude, a range far greater than for the H~II regions and stars and greater than can be explained by the uncertainties in abundance determinations. In addition, carbon levels in PNe appear on average greater than those typical of H~II regions for the same oxygen value, indicating that additional carbon above the level present in the interstellar medium at the time the star formed was produced during the progenitor stars' lifetimes.

Similar comments may be made concerning nitrogen abundances displayed in Fig.~3B. Here, while H~II regions seem to suggest a relation between nitrogen and metallicity as gauged by oxygen in the interstellar medium, we see no such pattern for PNe. And, as in the case of carbon, nitrogen abundances for PNe tend systematically to be greater than those for H~II regions, again suggesting that nitrogen is produced by PN progenitor stars.

Support for the contention that Figs.~3A and B imply that only C and N are synthesized in IMS is strengthened by comparing our PN abundances for Ne and O with those found in H~II regions. These two elements are apparently formed together in massive stars, and so their levels are expected to climb in lockstep as a galactic system evolves. A plot of interstellar log(Ne/O) versus 12+log(O/H) is therefore expected to show a constant value for Ne/O over a range in O abundance. Henry \& Worthey (1999) recently compiled abundance data for H~II regions and their results are shown with open circles in Fig.~3C. The flat behavior of Ne/O particularly at low levels of O/H is clearly seen. Scatter is undoubtedly larger at higher O/H levels due to lower electron temperatures at higher metallicity and the greater uncertainty in measuring the temperature. 
There is little expectation that PN progenitors will alter the levels of Ne and O that were initially present in the stars at the time of formation. Thus,
PN abundances of Ne and O are expected to reflect levels in the interstellar medium at the time of progenitor formation and should exhibit an interstellar pattern. Our PNe are shown with filled circles, and one can see that the pattern displayed by them is indistinguishable from that of the H~II regions. The two exceptions marked in the figure are the halo PNe BB1 and H4-1. Their unusual Ne/O ratios perhaps indicate that the halo material out of which they formed was poorly mixed. These objects need to be investigated further.
We see, therefore, that observed C and N abundance patterns in PNe suggest that abundances of these two elements are altered by processes related to the evolution of the progenitor star. 

We next use our PN abundance results to test the theoretical predictions of PN abundances. The real motivation here is to evaluate yield predictions of IMS, since it is these predictions which are used to study the origins of carbon and nitrogen using chemical evolution models.

We test two published sets of theoretical calculations. van~den~Hoek \& Groenewegen (1997; VG) calculated a grid of stellar models ranging in mass fraction metallicity between 0.001 and 0.04 and progenitor mass of 0.8 to 8~M$_{\odot}$. Likewise, Marigo, Bressan, \& Chiosi (1996; MBC) calculated models for
mass fraction metallicity of 0.008 and 0.02 for stars between 0.7 and 5~M$_{\odot}$. Both teams
employed up-to-date information about opacities and mass loss to calculate yields for several isotopes, including $^{4}$He, $^{12}$C, $^{13}$C, and $^{14}$N. But while it is the total yields which we wish to validate, it is really the PN abundances which can be observed directly.
Thus, both teams predicted PN abundances as a function of progenitor mass and metallicity, and it is these values we wish to compare with our PN abundances.

In the case of the VG calculations, we used information in their tables headed ``Final AGB yields'' along with their adopted progenitor abundances and their equation~4 to compute predicted PN abundances by number and abundance ratios for comparison with our observed abundances. MBC tabulated the abundances directly, and so no conversion was necessary for their predictions. 
The results are shown in Figs.~4 and 5.

Figs.~4 are linear plots of abundance pairs, with each abundance normalized to its solar value (Grevesse et al. 1996). Our PN abundances are shown with filled circles, the VG model predictions are shown with solid lines, and those of MBC are shown with dot-dashed lines. Each line connects models representing a constant progenitor mass over a range in metallicity, where the mass is indicated with integers along a line.

Fig.~4A is a plot of C versus O. Note first that at all metallicities the carbon abundance is initially predicted to rise with mass but then drop back to low values as mass continues to increase above 2-3~M$_{\odot}$. This reversal is the result of hot-bottom burning in stars with masses exceeding this level in which carbon from the 3rd dredge-up is converted to nitrogen at the base of the convective envelope late in the AGB stage. As a result, we can see in Fig.~4A that the observed abundances are consistent with predictions for both high and low mass progenitors.
 
Fig.~4B is similar to Fig.~4A but for nitrogen.
We do not include the tracks for stars above 4~M$_{\odot}$, as they are out of range of our PN abundances. We see that here the predicted behavior of nitrogen with progenitor mass is positively monotonic and seems to be consistent with the observed trend. At the same time, we see that the nitrogen abundances are explained by relatively low mass progenitors, and so this helps remove the mass ambiguity seen in Fig.~4A, i.e. apparently the C and N abundances observed in PNe are consistent with progenitor masses in the range of 1-4~M$_{\odot}$.

Fig.~4C reinforces our conclusions about the progenitor mass range. Here we plot N versus He. Again we see that theoretical predictions for progenitors in the 1-4~M$_{\odot}$ range are consistent with observations. The one extreme outlier is PB6, whose unusually high helium abundance needs to be confirmed with an independent set of observations.

Figs.~5A,B,C show PN abundances of C, N, and He, respectively, plotted against progenitor mass, where our sample object remnant masses were taken from the recent results of G{\'o}rny et al. (1997) and Stasi{\'n}ska et al. (1997) and converted to progenitor masses using the initial-final mass relation of Weidemann (1987). Solid (VG) and dashed (MBC) lines now connect model predictions for a constant metallicity over a range in progenitor mass. The value of the metallicity associated with each model line is indicated. 

The PN carbon abundances in Fig.~5A fall systematically below the predictions, while the nitrogen abundances in Fig.~5B compare slightly more favorably with theory, particularly the predictions of MBC, as do the helium abundances in Fig.~5C. But in Figs. 5B and 5C observed abundances appear to be more consistent with models representing metallicities below those expected for their given masses and evolutionary times. Data in Figs.~5A and 5B are consistent with the predicted complementary behavior of carbon and nitrogen in which carbon rises initially with mass until hot-bottom burning sets in, at which time carbon falls as carbon from the 3rd dredge-up is converted to nitrogen. However, due to the lack of progenitors in the important 3.5-6.5M$_{\sun}$ range, we are unable to test the predictions further. 

There are several obvious problems which could explain the less than satisfactory agreement in Figs.~5. First, we must keep in mind that the progenitor masses determined for our objects are very suspect, since in each case we take an uncertain remnant mass and convert it to a progenitor mass using an uncertain initial-final mass relation. Clearly, points at 0.5~M$_{\sun}$ are incorrect, since progenitors with this mass could not have produced a PN yet. Second, the PN carbon and nitrogen abundances in Figs.~5A,B may be systematically several times below their actual levels if temperature fluctuations (see {\S}4.1) are important, and so our C and N abundances may actually be lower limits. Third, the scatter in the observed PN abundances may make it difficult to constrain the models sufficiently. Given current observational and abundance-determination techniques, it is hard to obtain results with scatter of less than $\pm$50\%. And finally, the models themselves could be partially to blame for the lack of good agreement, since they rely heavily on mixing and mass loss laws which are currently very difficult to constrain.   

In summary, if observational confirmation of predicted PN abundances can be extended to the predicted stellar yields, then
our comparisons in Figs. 4 and 5
provide some modest empirical support for the latter. Furthermore,
we believe this rough agreement justifies confidence in the predicted
IMS stellar yields to levels of better than the order-of-magnitude, but
probably not better than within factors of 2 or 3. It is imperative,
however, that the models continue to be tested with larger samples of
PNe whose abundances have been carefully determined. Likewise, a better
understanding of the initial-final mass relation for the progenitor
stars and laws of mass loss will go a long way to improve model
calculations. And as improvements in both theory and observation are
made, we will be better able to ascertain the exact role that
intermediate mass stars play in the synthesis of carbon, nitrogen, and
helium in galaxies.

\section{Project Summary}

We began this project in 1993 with the goal of using the newly released Final Archive UV data from the IUE along with optical spectrophotometry to derive the most accurate carbon abundances possible for a well-defined and moderate-sized sample of Galactic planetary nebulae spanning a wide range in progenitor mass and metallicity. We made a careful attempt to choose optical data for positions within each object which overlapped as closely as possible the slit positions of the IUE. To ensure this in many cases we obtained our own data. Observed line strengths and the results of the abundance determinations have been documented by previous progress reports (Papers I, II, and III, and a dedicated paper on NGC~7293) plus the first part of the current paper. We have developed a technique for determining abundances using photoionization models to match a set of diagnostic line ratios to apply a final correction to the standard abundance method. This correction adjusts for the effects of such processes as charge exchange and dielectronic recombination which can alter the ionization structure of a nebula and otherwise lessen the accuracy of the ionization correction factors when the latter are used by themselves.
We began by focusing on carbon, but 
the project quickly expanded into a consideration of other elements, in particular nitrogen, since carbon and nitrogen abundances in galaxies are both apparently affected by nucleosynthesis in intermediate mass stars. In the end, what we feel we have achieved is an accurate set of abundances for a sample of well-known planetary nebulae whose progenitor stars span a relatively broad range in mass and metallicity.  

A second project aim was to use our abundances as constraints on published stellar model predictions of PN abundances in order to assess the applicability of associated stellar yields to the further study of galactic chemical evolution. In {\S}4 we explored this issue in some detail using the PN abundance predictions of van~den~Hoek \& Groenewegen (1997) and Marigo et al. (1996).

Our major conclusions from the project are:

\begin{itemize}

\item Carbon and nitrogen abundances in planetary nebulae, when plotted against oxygen, show a much broader range
than H~II regions and F and G stars and are generally higher. At the same time, both oxygen and neon display similar patterns in both PNe and H~II regions. Taken together, these results support the idea that PN progenitors synthesize significant amounts of carbon and nitrogen.

\item  Abundances of carbon, nitrogen, and helium found in our sample of PNe show some consistency with model predictions. We believe that this is the first time that such a detailed comparison of observation and theory has been possible, and that our results provide modest support for the use of published yields of intermediate mass stars in studying galactic chemical evolution, especially in the cases of carbon and nitrogen. We further believe that observation and theory for PN abundances currently agree to well within an order of magnitude but probably not better than a factor of 2 or 3; this same level of confidence would then also apply to the predicted yields.

\item Our comparisons of observed and predicted PN abundances modestly support the predicted occurrence of hot-bottom burning in stars above about 3.5-4~M$_{\odot}$, but a gap in progenitor mass for the observed sample between 3.5-=6.5M$_{\sun}$ prevents us from confirming hot-bottom burning completely from the PN abundances.

\end{itemize}

Further tests of intermediate mass star nucleosynthesis should be undertaken with a larger sample to improve statistics and to confirm the suggestions of our comparison here. It is extremely important in the study of carbon and nitrogen evolution that we understand the contributions made by intermediate mass stars and how these contributions compare with those of massive stars which erupt in supernovae. But in summary, we are encouraged for now by the apparent consistency between observed and theoretical PN abundances and how this consistency reflects positively on the stellar yield predictions offered by the same calculations. 

There is also a need to reconfirm and understand certain examples of anomalous abundances. For example, the two halo PNe BB1 and H4-1 both deviate from other PNe in terms of their Ne/O ratio by a significant amount, suggesting that these objects are revealing interesting conditions in the halo at the time their progenitors were formed. It would be desirable to verify the neon abundances in each of these objects by using additional UV spectra to measure another ionization stage, i.e. Ne$^{+4}$, since currently only the Ne$^{+2}$ stage is usually used to determine the abundance of this element.

\acknowledgments

We are grateful to the support staff at KPNO for help in carrying out the observing portions of this program.
This project was supported by NASA grant NAG 5-2389.


\begin{deluxetable}{lccc}
\tablecolumns{4}
\tablewidth{0pc}
\tablenum{1A}
\tablecaption{Optical Spectra}
\tablehead{
\colhead{Object} &
\colhead{Slit Position\tablenotemark{1}} &
\colhead{Blue Exp (sec)}&
\colhead{Red Exp (sec)}
} 
\startdata 
IC 418 & 5$\arcsec$N  & 35 & 440  \\
NGC 2392 A & 14$\arcsec$S & 450 & 2520 \\
NGC 2392 B & 14$\arcsec$N & 390 & 1500 \\
NGC 2392 C & on central star & 300 & 900 \\
NGC 2392 D & on central star & 300 & 900 \\
NGC 3242 & 8$\arcsec$S  & 130  & 120 \\
\enddata
\tablenotetext{1}{Positions are for slit center relative to the central star.}
\end{deluxetable}

\begin{deluxetable}{lcccc}
\tablecolumns{5}
\tablewidth{0pc}
\tablenum{1B}
\tablecaption{IUE Final Archive Spectra}
\tablehead{
\colhead{Object} &
\colhead{SWP} &
\colhead{Offset\tablenotemark{1}} &
\colhead{Slit PA ($^\circ$)} &
\colhead{Exp (sec)}
} 
\startdata 
IC 418 & 08235 & 2$\arcsec$N, 7$\arcsec$E  & 341 & 600 \\
IC 418 & 08236 & 2$\arcsec$N, 7$\arcsec$E  & 341 & 2099 \\
NGC 2392 A & 45786 & 15$\arcsec$S  & 216 & 420 \\
NGC 2392 A & 45787 & 15$\arcsec$S	 & 171  & 1800 \\
NGC 2392 B & 45788 & 15$\arcsec$N & 171 & 1800 \\
NGC 2392 C & 45789 & 8$\arcsec$E & 171 & 600 \\
NGC 2392 C & 45790 & 8$\arcsec$E & 171 & 1200 \\
NGC 2392 D & 45791 & 8$\arcsec$W & 171  & 1200 \\
NGC 3242 & 16418 & 8S$\arcsec$ & 257 & 2700 \\
NGC 3242 & 16419 & 8S$\arcsec$ & 257 & 180 \\
\enddata 
\tablenotetext{1}{Offsets are for slit center relative to the central star.}
\end{deluxetable}

\begin{deluxetable}{lrrrccccc}
\tiny
\tablecolumns{9}
\tablewidth{0pc}
\tablenum{2A}
\tablecaption{UV \& Optical Line Strengths}
\tablehead{
\colhead{} & \colhead{} &
\multicolumn{2}{c}{IC 418} &
\colhead{NGC 2392A\tablenotemark{a}} &
\colhead{NGC 2392B\tablenotemark{a}} &
\colhead{NGC 2392C\tablenotemark{a}} &
\colhead{NGC 2392D\tablenotemark{a}} &
\colhead{NGC 3242\tablenotemark{a}} \nl
\cline{3-4} \cline{5-9} \nl
\colhead{Line} &
\colhead{f($\lambda$)} &
\colhead{F($\lambda$)} &
\colhead{I($\lambda$)} &
\colhead{I($\lambda$)} &
\colhead{I($\lambda$)} &
\colhead{I($\lambda$)} &
\colhead{I($\lambda$)} &
\colhead{I($\lambda$)}
}
\startdata
C III $\lambda$1175 &1.85&\nodata&\nodata&\nodata&\nodata&\nodata&\nodata&29 \nl
N V $\lambda$1241 &1.64&\nodata&\nodata&\nodata&\nodata&\nodata&\nodata&5 \nl
C II $\lambda$1336 &1.41&16&15&5&15:&10::&10::&11 \nl
O IV] $\lambda$1402 &1.31&\nodata&\nodata&12&19:&26::&29:&7: \nl
N IV] $\lambda$1485 &1.23&\nodata&\nodata&11&16:&26:&22:&7 \nl
C IV $\lambda$1549 &1.18&\nodata&\nodata&36&71&71&78&36 \nl
\[[Ne V] $\lambda$1575 &1.17&\nodata&\nodata&\nodata&\nodata&\nodata&\nodata&7: \nl
He II $\lambda$1640 &1.14&\nodata&\nodata&157&235&256&241&166 \nl
O III] $\lambda$1662 &1.13&\nodata&\nodata&31&56&65&58&17 \nl
N III] $\lambda$1750 &1.12&\nodata&\nodata&24&57&59&54&8 \nl
C II $\lambda$1760 &1.12&3&2&\nodata&\nodata&\nodata&\nodata&\nodata \nl
Si III] $\lambda$1887 &1.21&\nodata&\nodata&12&31&36::&29&\nodata \nl
C III] $\lambda$1909 &1.23&32&28&\nodata&183&207&204&235 \nl
\[[O II] $\lambda$ 3727 &0.29&105&115&122&107&95&122&9 \nl
He II + H10 $\lambda$3797 & 0.27&3&4&4&4&\nodata&\nodata&4 \nl
He II + H9 $\lambda$3835&0.25&7&8&8&9&\nodata&\nodata&9 \nl
\[[Ne III] $\lambda$3869 &0.25&2&2&93&131&96&92&119 \nl
He I + H8 $\lambda$3889 &0.25&11&12&18&18&\nodata&\nodata&16 \nl
H$\epsilon$ + \[[Ne III] $\lambda$3968 &0.23&21&22&63&76&129&135&72 \nl
He II $\lambda$4026 &0.21&1:&1:&1:&1:&\nodata&\nodata&1: \nl
\[[S II] $\lambda$4072 &0.20&3:&3:&3:&2:&6::&\nodata&1: \nl
He II + H$\delta$ $\lambda$4101&0.19&18&19&22&23&24&27&24 \nl
He II $\lambda$4198 &0.16&\nodata&\nodata&0.4:&0.5:&\nodata&\nodata&0.3: \nl
C II $\lambda$4267 &0.14&0.3:&0.3:&\nodata&\nodata&\nodata&\nodata&0.6: \nl
H$\gamma$ $\lambda$4340 &0.13&40&41&45&46&45&47&44 \nl
\[[O III] $\lambda$4363 &0.12&0.5::&0.5::&13&20&12&10&11 \nl
He I $\lambda$4471 &0.09&3&3&3&2&\nodata&\nodata&3 \nl
He II $\lambda$4540 &0.07&\nodata&\nodata&0.8:&0.9:&\nodata&\nodata&0.7: \nl
N III $\lambda$4640 &0.05&\nodata&\nodata&0.3:&0.3:&10&11&4 \nl
He II $\lambda$4686 &0.04&\nodata&\nodata&23&34&38&36&25 \nl
\[[Ar IV] + He I $\lambda$4712 &0.03&0.4::&0.4::&3&3&\nodata&\nodata& 4\nl
\[[Ar IV] $\lambda$4740 &0.02&\nodata&\nodata&1:&2&\nodata&\nodata& 4\nl
H$\beta$ $\lambda$4861 &0.00&100&100&100&100&100&100&100 \nl
He I $\lambda$4922 &-0.02&1:&1:&0.7:&0.6:&3&\nodata&0.9: \nl
\[[O III] $\lambda$4959 &-0.03&37&36&264&343&289&292&339 \nl
\[[O III] $\lambda$5007 &-0.04&125&123&845&1125&911&822&1107 \nl
\[[N I] $\lambda$5199 &-0.09&0.5:&0.5:&2&0.7:&\nodata&\nodata&\nodata \nl
He II $\lambda$5411 &-0.13&\nodata&\nodata&2&3&\nodata&\nodata&2 \nl
\[[Cl III] $\lambda$5517 &-0.16&0.2:&0.2:&0.5:&0.6:&\nodata&\nodata&0.2: \nl
\[[N II] $\lambda$5755 &-0.21&4&4&2&2&2&2:&\nodata \nl
He I $\lambda$5876 &-0.23&11&10&8&6&5&6&9 \nl
\[[K IV] $\lambda$6101 & -0.28&\nodata&\nodata&\nodata&\nodata&\nodata&\nodata& 0.2:\nl
\[[O I] $\lambda$6300 &-0.31&4&4&5&3&0.4:&0.3:&\nodata \nl
\[[S III] $\lambda$6312 &-0.31&1::&1::&3&3&2&2&0.6: \nl
\[[O I] $\lambda$6363 &-0.32&1::&1::&1::&0.8::&\nodata&\nodata&\nodata \nl
\[[N II] $\lambda$6548 &-0.36&66&59&38&30&26&25&0.6:: \nl
H$\alpha$ $\lambda$6563 &-0.36&321&286&283&270&279&260&285 \nl
\[[N II] $\lambda$6584 &-0.36&197&175&112&87&74&75&2 \nl
He I $\lambda$6678 &-0.38&3&3&2&2&2&2&2 \nl
\[[S II] $\lambda$6716 &-0.39&3&3&14&9&5&5&0.2: \nl
\[[S II] $\lambda$6731 &-0.39&6&5&15&10&7&7&0.3: \nl
\[[Ar V] $\lambda$7005 &-0.43&0.3:&0.2:&\nodata&\nodata&\nodata&\nodata&\nodata \nl
He I $\lambda$7065 &-0.44&8:&7:&2:&2:&3:&3:&3: \nl
\[[Ar III] $\lambda$7135 &-0.45&9:&8:&13:&15:&13:&13:&7: \nl
\[[Ar IV] $\lambda$7168 &-0.46&\nodata&\nodata&\nodata&\nodata&\nodata&\nodata& 0.2:\nl
He II $\lambda$7178 &-0.46&\nodata&\nodata&0.2:&0.5:&\nodata&\nodata&\nodata \nl
\[[Ar IV] $\lambda$7235 &-0.47&0.6:&0.5:&0.1:&0.1:&\nodata&\nodata&0.2: \nl
He I $\lambda$7281 &-0.47&0.9:&0.8:&0.4:&0.4:&0.9:&0.8:&0.4: \nl
\[[O II] $\lambda$7325 &-0.48&42:&36:&4:&5:&9:&7:&1: \nl
\[[Cl IV] $\lambda$7529 &-0.51&\nodata&\nodata&0.1:&0.2:&\nodata&\nodata&0.3: \nl
\[[Ar III] $\lambda$7751 &-0.54&2:&2:&3:&4:&4:&3:&2: \nl
\[[Cl IV] $\lambda$8045 &-0.57&\nodata&\nodata&0.3:&0.5::&\nodata&\nodata&0.7: \nl
He II $\lambda$8236 &-0.59&\nodata&\nodata&0.6:&0.9::&\nodata&\nodata&0.6: \nl
P16 $\lambda$8467 &-0.62&0.8:&0.6:&0.4:&0.4::&\nodata&\nodata&0.3: \nl
P15 $\lambda$8502 &-0.62&1:&0.8:&0.5:&0.5::&\nodata&\nodata&0.4: \nl
P14 $\lambda$8544 &-0.63&1:&0.8:&0.5:&0.6::&\nodata&1::&0.5: \nl
P13 $\lambda$8598 &-0.63&1:&1:&0.6:&0.7::&\nodata&\nodata&0.6: \nl
P12 $\lambda$8664 &-0.64&2:&1:&0.9:&1::&2::&2::&0.9: \nl
P11 $\lambda$8750 &-0.64&2:&2:&1:&1::&2::&2::&1: \nl
P10 $\lambda$8862 &-0.65&3:&2:&1:&1::&1::&1::&1: \nl
P9 $\lambda$9014 &-0.67&0.9:&0.7:&1:&1::&\nodata&1::&2: \nl
\[[S III] $\lambda$9069 &-0.67&34:&27:&23:&30::&31::&27::&7: \nl
P8 $\lambda$9228 &-0.68&6:&5:&3:&3::&4::&4::&3: \nl
\[[S III] $\lambda$9532 &-0.70&102:&81:&75:&99::&95::&87::&21: \nl
P7 $\lambda$9544 &-0.70&10:&8:&5:&6::&4::&4::&3: \nl\nl
\samepage
\nopagebreak
log F$_{H\beta}$\tablenotemark{b}&&-10.50&&-11.27&-11.60&-11.23&-11.12&-10.48 \nl
c&&0.14&&0&0&0&0&0 \nl
merging factor\tablenotemark{c}&&0.58&&0.54&0.61&0.88&1.14&1.20 \nl
\nopagebreak
\enddata
\samepage
\nopagebreak
\tablenotetext{a}{This object is not reddened, so final intensities are equal to observed fluxes.}
\tablenotetext{b}{Ergs/cm$^2$/s in our extracted spectra}
\tablenotetext{c}{Factor by which dereddened UV line strengths are multiplied in order to merge them with optical data (see text).}
\nopagebreak
\end{deluxetable}

\clearpage

\begin{deluxetable}{lccccccc}
\tiny
\tablecolumns{8}
\tablewidth{0pc}
\tablenum{2B}
\tablecaption{Line Ratios}
\tablehead{
\colhead{} & \colhead{} & \multicolumn{6}{c}{Observed}\\ \cline{3-8} \nl
\colhead{Ratio} & 
\colhead{Theory} & 
\colhead{IC 418} & 
\colhead{NGC 2392A} & 
\colhead{NGC 2392B} &
\colhead{NGC 2392C} &
\colhead{NGC 2392D} &
\colhead{NGC 3242} 
}
\startdata
[Ne~III] 3869/3968\tablenotemark{a} & 3.32 &0.30&1.95&2.16&0.84&0.77&2.10 \nl
He~I 5876/4471 & 2.76 &3.33&2.67&3.00&\nodata&\nodata&3.00 \nl
[O~III] 5007/4959\tablenotemark{b} & 2.89 &3.42&3.20&3.28&3.15&2.82&3.27 \nl
[N~II] 6584/6548 & 2.95 &2.97&2.95&2.90&2.85&3.00&3.33 \nl
He~I 6678/4471 & 0.79 &1.00&0.67&1.00&\nodata&\nodata&0.67 \nl
[Ar~III] 7135/7751 & 4.14 &4.00&4.33&3.75&3.25&4.33&3.50 \nl
P8/H$\beta$ 9228/4861 & 0.037 &0.05&0.03&0.03&0.04&0.04&0.03 \nl
[S~III] 9532/9069 & 2.48 &3.00&3.26&3.30&3.06&3.22&3.00 \nl
\enddata
\tablenotetext{a}{The [Ne~III] $\lambda$3968 line was corrected for the contribution from H$\epsilon$.}
\tablenotetext{b}{The observed line ratios are somewhat unreliable, since the $\lambda$4959 line often fell along a bad column.}
\end{deluxetable}

\clearpage

\begin{deluxetable}{lllllll}
\tiny
\tablecolumns{7}
\tablewidth{0pc}
\tablenum{3}
\tablecaption{Ion Abundances \& Ionization Correction Factors}
\tablehead{
\colhead{Ion Ratio} &
\colhead{IC 418} &
\colhead{NGC 2392A} &
\colhead{NGC 2392B} &
\colhead{NGC 2392C} &
\colhead{NGC 2392D} &
\colhead{NGC 3242} 
}
\startdata
He$^+$/H$^+$ &0.069&0.058&0.042&0.035&0.042&0.063 \nl
He$^{+2}$/H$^+$ &\nodata&0.021&0.031&0.035&0.033&0.023  \nl
ICF(He) &1.00&1.00&1.00&1.00&1.00&1.00 \nl\nl
O$^+$/H$^+$($\times 10^{4}$) &0.62&0.48&0.28&0.26&0.35&0.041  \nl
O$^{+2}$/H$^+$($\times 10^{4}$) &0.77&1.39&1.56&1.77&1.75&2.91  \nl
ICF(O) &1.00&1.36&1.74&2.01&1.79&1.36  \nl\nl
C$^{+2}$/H$^+$($\times 10^{4}$) &2.79&\nodata&0.51&1.24&1.49&2.82 \nl
C$^{+3}$/H$^+$($\times 10^{5}$) &\nodata&1.08&1.35&3.44&4.75&4.04 \nl
ICF(C) &2.23&5.30&11.4&15.7&10.7&98.7  \nl\nl
N$^+$/H$^+$($\times 10^{5}$) &3.69&1.88&1.14&1.01&1.06&0.036  \nl
ICF(N) &2.23&5.30&11.4&15.7&10.7&98.7  \nl\nl
Ne$^{+2}$/H$^+$($\times 10^{5}$) &0.42&3.66&4.20&4.67&4.97&8.36  \nl
ICF(Ne) &1.81&1.84&2.05&2.31&2.15&1.38  \nl\nl
\enddata
\end{deluxetable}

\clearpage

\begin{deluxetable}{lrrrrrrrrrrrr}
\tiny
\tablecolumns{13}
\tablewidth{0pc}
\tablenum{4A}
\tablecaption{Observations \& Models}
\tablehead{
\colhead{} & 
\multicolumn{2}{c}{IC 418} &
\multicolumn{2}{c}{NGC 2392A} &
\multicolumn{2}{c}{NGC 2392B} &
\multicolumn{2}{c}{NGC 2392C} &
\multicolumn{2}{c}{NGC 2392D} &
\multicolumn{2}{c}{NGC 3242} \nl
\colhead{} &
\colhead{Obs} &
\colhead{Model} &
\colhead{Obs} &
\colhead{Model} &
\colhead{Obs} &
\colhead{Model} &
\colhead{Obs} &
\colhead{Model} &
\colhead{Obs} &
\colhead{Model} &
\colhead{Obs} &
\colhead{Model}
}
\startdata
log ($I_{[O II]}+I_{[O III]}$)/H$\beta$ &0.45&0.42&1.10&1.07&1.21&1.20&1.12&1.21&1.09&1.23&1.18&1.11 \nl
log $I_{[O II]}/I_{[O III]}$ &-0.16&0.02&-0.97&-1.00&-1.15&-1.13&-1.11&-1.11&-0.96&-1.13&-2.22&-2.31 \nl
log I$_{He II}$/I$_{He I}$ &\nodata&-3.06&0.46&0.50&0.75&0.79&0.88&0.68&0.78&0.60&0.44&-0.36 \nl
log I$_{\lambda 4363}$/I$_{\lambda 5007}$ &-2.39&-2.11&-1.81&-1.61&-1.75&-1.69&-1.88&-1.68&-1.91&-1.69&-2.00&-2.13 \nl
log I$_{\lambda 6716}$/I$_{\lambda 6731}$ &-0.22&-0.31&-0.03&-0.04&-0.05&-0.01&-0.15&-0.18&-0.15&-0.18&-0.18&-0.15 \nl
log I$_{He~I}$/I$_{H\beta}$ &-1.00&-0.93&-1.10&-1.10&-1.22&-1.23&-1.30&-1.15&-1.22&-1.12&-1.05&-0.86 \nl
log I$_{6584}$/I$_{3727}$ &0.18&0.20&-0.04&-0.22&-0.09&-0.09&-0.11&-0.03&-0.21&-0.28&-0.65&-0.66 \nl
log I$_{6724}$/I$_{3727}$ &-1.16&-1.04&-0.62&-0.71&-0.75&-0.67&-0.90&-0.98&-1.01&-0.99&-1.26&-1.25 \nl
log I$_{1909}$/I$_{5007}$ &-0.64&-0.55&\nodata&-1.50&-0.79&-0.68&-0.64&-0.62&-0.61&-0.64&-0.67&-0.71 \nl
log I$_{3869}$/I$_{5007}$ &-1.79&-1.73&-0.96&-0.95&-0.93&-0.97&-0.98&-0.95&-0.95&-0.95&-0.97&-1.03 \nl
\cutinhead{Model Input Parameters}
T$_{eff}$ (10$^3$K) &&38.0&&135&&148&&148&&140&&60 \nl
log L/L$_{\sun}$  &&3.3&&4.6&&5.8&&4.5&&4.5&&4.3 \nl
N$_e$ (cm$^{-3}$) &&12200&&1000&&870&&3400&&3400&&2000 \nl
R$_o$(pc) &&0.032&&0.032&&0.032&&0.032&&0.032&&0.032 \nl
R (pc) &&0.038&&0.50&&1.16&&0.20&&0.19&&0.11 \nl
He/H &&0.10&&0.08&&0.08&&0.08&&0.08&&0.10 \nl
O/H ($\times 10^4$) &&1.48&&1.11&&2.15&&2.15&&2.15&&2.96 \nl
C/O  &&0.72&&0.05&&0.48&&0.48&&0.48&&1.68 \nl
N/O  &&0.40&&0.38&&0.50&&0.50&&0.25&&0.15 \nl
Ne/O  &&0.068&&0.21&&0.21&&0.21&&0.21&&0.23 \nl
S/O  &&0.019&&0.02&&0.02&&0.01&&0.01&&0.01 \nl
Ar/O  &&0.008&&0.005&&0.005&&0.005&&0.005&&0.001 \nl
\enddata
\end{deluxetable}

\clearpage

\begin{deluxetable}{lcccccc}
\tablecolumns{7}
\tablewidth{0pc}
\tablenum{4B}
\tablecaption{Correction Factors ($\xi$)}
\tablehead{
\colhead{Ratio} &
\colhead{IC 418} &
\colhead{NGC 2392A} &
\colhead{NGC 2392B} &
\colhead{NGC 2392C} &
\colhead{NGC 2392D} & 
\colhead{NGC 3242} 
}
\startdata 
He/H&1.24&1.04&1.01&1.04&1.00&0.94 \nl
O/H&1.10&0.87&0.77&0.81&0.83&0.84 \nl
C/O&0.39&\nodata&1.37&1.22&1.21&1.26 \nl
N/O&0.95&1.01&1.24&1.00&1.01&2.05 \nl
Ne/O&1.16&0.85&0.88&0.84&0.83&0.84 \nl
\enddata
\end{deluxetable}

\clearpage

\begin{deluxetable}{lllllllll}
\tiny
\tablecolumns{9}
\tablewidth{0pc}
\tablenum{5}
\tablecaption{Derived Abundances, Temperatures, \& Densities}
\tablehead{
\colhead{Ratio} &
\colhead{IC 418} &
\colhead{NGC 2392A} & 
\colhead{NGC 2392B} & 
\colhead{NGC 2392C} & 
\colhead{NGC 2392D} &
\colhead{NGC 2392 (Ave)} & 
\colhead{NGC 3242} & 
\colhead{Sun\tablenotemark{a}} 
}
\startdata
He/H&0.086(.01)&0.082(.01)&0.073(.01)&0.072(.01)&0.075(.01)&0.076(.01)&0.081(.01)&0.10 \nl
O/H($\times 10^{4}$)&1.53(.23)&2.22(.33)&2.46(.37)&3.32(.50)&3.12(.47)&2.78(.42)&3.38(.51)&7.41 \nl
C/O&1.43(.57)&\nodata&0.45(.18)&0.85(.34)&1.03(.41)&0.78(.31)&1.22(.49)&0.48 \nl
N/O&0.56(.22)&0.39(.16)&0.50(.20)&0.39(.16)&0.31(.12)&0.40(.16)&0.18(.07)&0.13 \nl
Ne/O&0.063(.01)&0.22(.03)&0.24(.04)&0.22(.03)&0.24(.04)&0.23(.03)&0.24(.04)&0.16 \nl
T$_{[O~III]}$($10^{3}$K)&8.5&13.0&13.8&12.2&11.9&12.7&11.1&\nodata \nl
T$_{[N~II]}$($10^{3}$K)&10.7&9.6&10.7&11.5&11.4&10.8&10.3&\nodata \nl
T$_{[O~II]}$($10^{3}$K)&17.1&5.6&6.3&8.4&6.8&6.8&9.0&\nodata \nl
T$_{[S~II]}$($10^{3}$K)&8.7&9.2&9.3&\nodata&11.4&10.0&\nodata&\nodata \nl
T$_{[S~III]}$($10^{3}$K)&8.1&13.9&11.9&10.0&10.4&11.6&11.5&\nodata \nl
N$_{e,[S II]}$(cm$^{-3}$)&3300&800&1000&2000&2000&2500&1450&\nodata \nl
\enddata
\tablenotetext{a}{Grevesse, Noels, \& Sauval (1996)}
\end{deluxetable}

\clearpage

\begin{center}
\begin{deluxetable}{lcccccccccc}
\tiny
\tablecolumns{11}
\tablewidth{0pc}
\tablenum{6}
\tablecaption{Derived Abundances, Temperatures, \& Densities For Complete Sample}
\tablehead{
\colhead{Object} &
\colhead{Type\tablenotemark{a}} &
\colhead{He/H} & 
\colhead{O/H(x10$^4$)} & 
\colhead{C/O} & 
\colhead{N/O} & 
\colhead{Ne/O} & 
\colhead{T$_{[O~III]}$(10$^3$K)} & 
\colhead{T$_{[N~II]}$(10$^3$K)} & 
\colhead{N$_e$(cm$^{-3}$)} & 
\colhead{Paper\tablenotemark{b}} 
}
\startdata
BB1&4&0.10(.02)&1.07(.15)&23.1(12)&1.11(.17)&0.81(.12)&12.0 &9.5 &1500 \tablenotemark{c}&2 \nl
DDDM1&4&0.10(.02)&1.17(.18)&0.05(.02)&0.30(.12)&0.12(.02)&11.9 &12.0 &3300 &3 \nl
Hu 2-1&2.5&0.11(.02)&3.11(.31)&1.79(.72)&0.38(.04)&0.11(.01)&9.0 &14.6 &17000 &1 \nl
H4-1&4&0.10(.01)&2.15(.20)&1.91(.29)&0.23(.03)&0.01(.002)&12.4 &11.5 &100 &1 \nl
IC 418&1&0.09(.01)&1.53(.23)&1.43(.57)&0.56(.22)&0.06(.01)&8.5 &10.7 &3300 &4 \nl
IC 3568&2.5&0.11(.02)&3.14(.47)&0.31(.12)&0.75(.30)&0.13(.02)&10.5 &6.9 &900 &3 \nl
IC 4593&3&0.10(.02)&5.95(.89)&0.08(.03)&0.17(.07)&0.11(.02)&8.2 &7.7 &1700 &3 \nl
K 648&4&0.08(.02)&0.41(.08)&4.68(1.9)&0.11(.03)&0.07(.02)&11.4 &9.2 &1000 &1 \nl
NGC 0650&1&0.13(.02)&5.61(.84)&2.96(1.48)&0.42(.06)&0.19(.03)&12.0 &9.1 &400 &2 \nl
NGC 1535&1.5&0.09(.01)&3.12(.47)&0.61(.06)&0.06(.01)&0.13(.02)&11.8 &\nodata&6000 \tablenotemark{d}&2 \nl
NGC 2392&1&0.08(.01)&2.78(.42)&0.78(.31)&0.40(.16)&0.23(.03)&12.7 &10.8 &2500 &4 \nl
NGC 2440&1&0.12(.02)&5.71(.86)&1.11(.22)&1.44(.22)&0.13(.02)&14.0 &9.6 &3300 &2 \nl
NGC 3242&2.5&0.08(.01)&3.38(.51)&1.22(.49)&0.18(.07)&0.24(.04)&11.1 &10.3 &1500 &4 \nl
NGC 6210&2.5&0.11(.02)&4.40(.66)&0.21(.08)&0.33(.13)&0.16(.02)&9.4 &10.9 &3200 &3 \nl
NGC 6720&2.5&0.11(.02)&5.94(.89)&1.09(.44)&0.42(.17)&0.17(.03)&10.3 &9.7 &400 &3 \nl
NGC 6826&2.5&0.10(.02)&3.88(.58)&0.34(.14)&0.23(.09)&0.13(.02)&8.9 &9.0 &1800 &3 \nl
NGC 7009&2.5&0.12(.02)&5.60(.84)&0.81(.32)&0.57(.23)&0.28(.04)&9.4 &10.7 &2700 &3 \nl
NGC 7027&2&0.11(.02)&5.08(.76)&1.88(.38)&0.32(.05)&0.27(.04)&14.3 &\nodata&128000 &2 \nl
NGC 7293&1&0.12(.02)&4.60(.18)&0.87(.12)&0.54(.14)&0.33(.04)&9.3 &9.7 &$<$100 &\nodata \nl 
PB6&1&0.20(\nodata)&6.29(\nodata)&2.67(\nodata)&0.43(\nodata)&0.18(\nodata)&15.1 &10.2 &2800 &1 \nl
Sun\tablenotemark{e}&\nodata&0.10&7.41&0.48&0.13&0.16&\nodata&\nodata&\nodata&\nodata \nl
Orion\tablenotemark{f}&\nodata&0.10&5.25&0.59&0.11&0.15&\nodata&\nodata&\nodata&\nodata
\enddata
\tablenotetext{a}{Type is defined by Peimbert (1978), where 1=I, 2=II, 3=III, 4=Halo, and intermediate types are indicated with decimal notation, e.g. 2.5=II-III.}
\tablenotetext{b}{Reference where details of the abundance determinations may be found. 1=Henry, Kwitter, \& Howard (1996); 2=Kwitter \& Henry (1996); 3=Kwitter \& Henry (1998); 4=this paper. NGC~7293 was treated separately in Henry, Kwitter, \& Dufour (1999).}
\tablenotetext{c}{From Torres-Peimbert, Rayo, \& Peimbert (1981)}
\tablenotetext{d}{From Guti{\'e}rrez-Moreno, Moreno, \& Cort{\'e}s (1986)}
\tablenotetext{e}{Grevesse, Noels, \& Sauval (1996)}
\tablenotetext{f}{Esteban et al. (1998), Table 19 (gas+dust)}
\end{deluxetable}
\end{center}

\newpage
\figurenum{1}\figcaption{Derived abundance ratios. Results from this study are shown with filled circles. Comparison abundances, shown with open circles, are from Hyung et al. (1994; IC~418) and Perinotto (1991; NGC~2392 and NGC~3242), with C/O ratios for NGC~2392 and NGC~3242 taken from Rola \& Stasi{\'n}ska (1994). Error bars are shown to represent typical uncertainties.}

\figurenum{2}\figcaption{Abundances determined in our series of studies (horizontal axis) plotted against results in the literature (vertical axis) for comparison. Sources for comparison abundances are as follows. BB1: Pe{\~n}a et al. (1991); DDDM1: Barker \& Cudworth (1984) and Howard et al. (1997); H4-1: Torres-Peimbert \& Peimbert (1979); IC~418: Hyung et al. (1994); IC4593: Bohigas \& Olgu{\'i}n (1996); K648: Adams et al. (1984); NGC~7293: Peimbert et al. (1995), Hawley (1978), and O'Dell (1998); PB6: Kaler et al. (1991); All remaining objects: all but C/O from Perinotto (1991); C/O from Rola \& Stasi{\'n}ska (1994) and Lutz (1981; Hu2-1 only). All of our C/O ratios and most comparison C/O values are derived from collisional lines of carbon; those C/O comparisons derived from carbon recombination lines are identified with an x inside the circle. Positions of BB1, IC~4593, and IC~3568 are identified in three panels and discussed briefly in the text. Error bars in each panel indicate typical uncertainties.}

\figurenum{3}\figcaption{A. log(C/O) versus 12+log(O/H) for our planetary nebulae (solid circles) and a combined sample of galactic and extragalactic H~II regions (open circles) discussed in Henry \& Worthey (1999), along with observations of Galactic B stars (open squares) by Gustafsson et al. (1999). The positions of the sun, M8, and and the Orion Nebula are indicated with S, M, and O, respectively. Representative error bars are shown. B. Same as A. but for log(N/O). C. Same as A. but for log(Ne/O). The locations of BB1 and H4-1, both halo PNe, are indicated.}

\figurenum{4}\figcaption{A. Carbon versus oxygen abundances, each normalized to its solar value (Grevesse et al. 1996), for our sample of planetary nebulae (filled circles). Solid lines show tracks for theoretical predictions from van den Hoek \& Groenewegen (1997), where each line connects predictions over a range in metallicity for the progenitor mass indicated at points along the line. Likewise, theoretical predictions from Marigo et al. (1996) are shown with dot-dashed lines. B. Same as A. but for nitrogen versus oxygen. C. Same as A. but for nitrogen versus helium.}

\figurenum{5}\figcaption{A. Carbon abundance normalized to the solar value (Grevesse et al. 1996) versus progenitor mass in solar units for our sample of planetary nebulae (filled circles). Masses are taken directly from G{\'o}rny, et al. (1997) and Stasi{\'n}ska et al. (1997). Also shown are theoretical predictions from van den Hoek \& Groenewegen (1997; solid lines) and Marigo et al. (1996; dashed lines) for the metallicities indicated for each line. B. Same as A. but for nitrogen. C. Same as A. but for helium.}

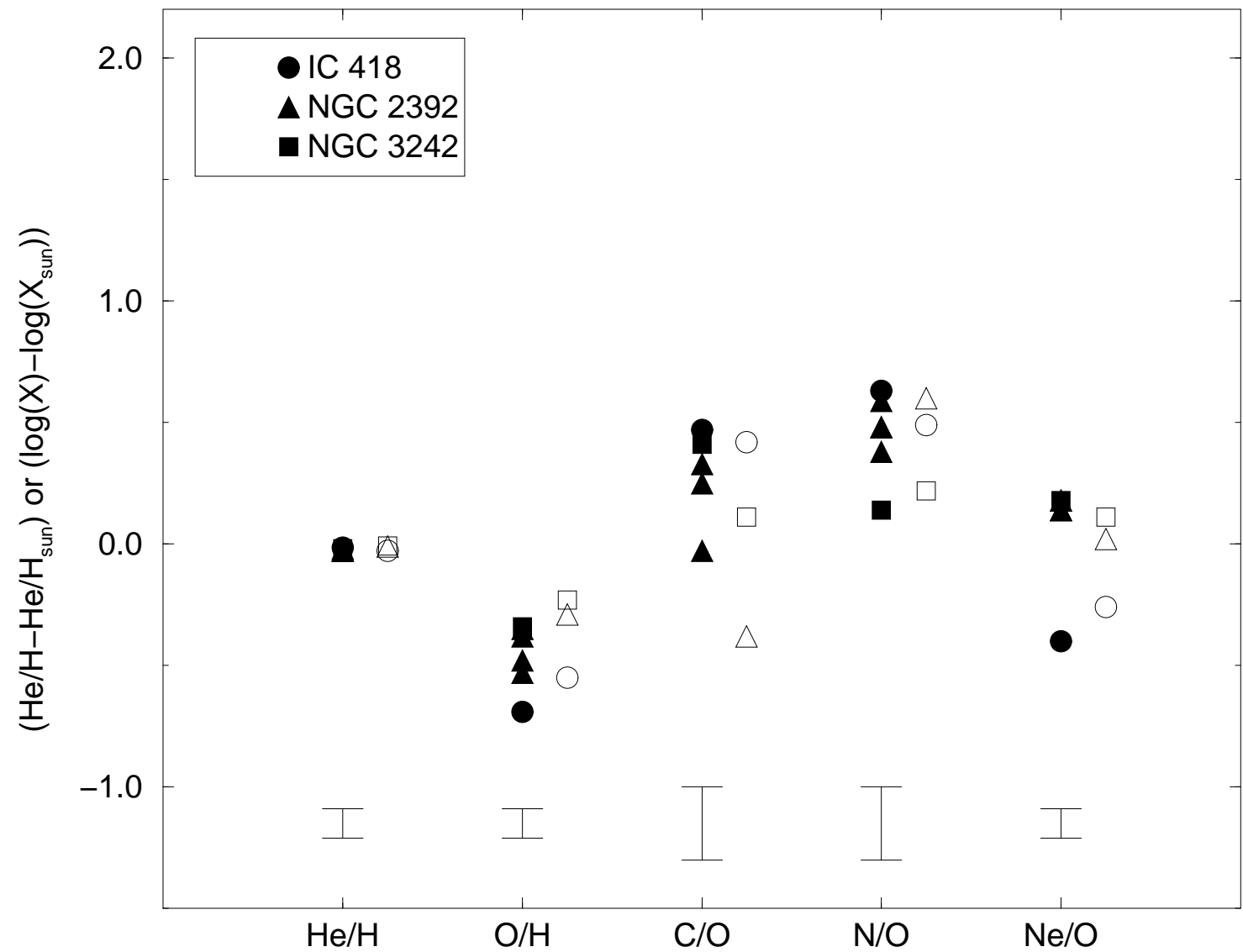
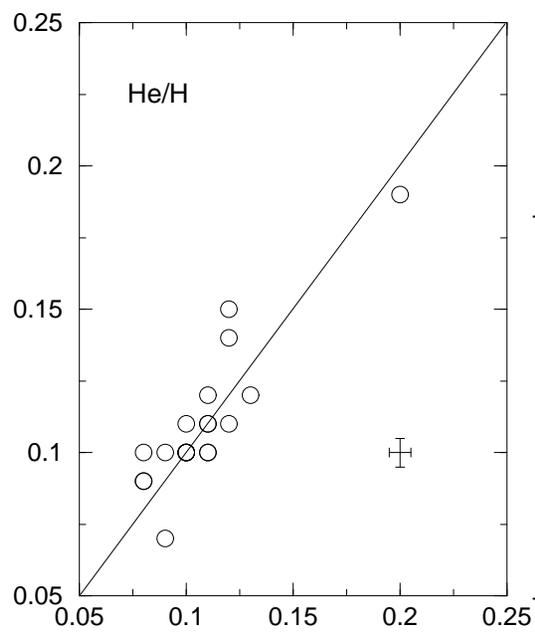
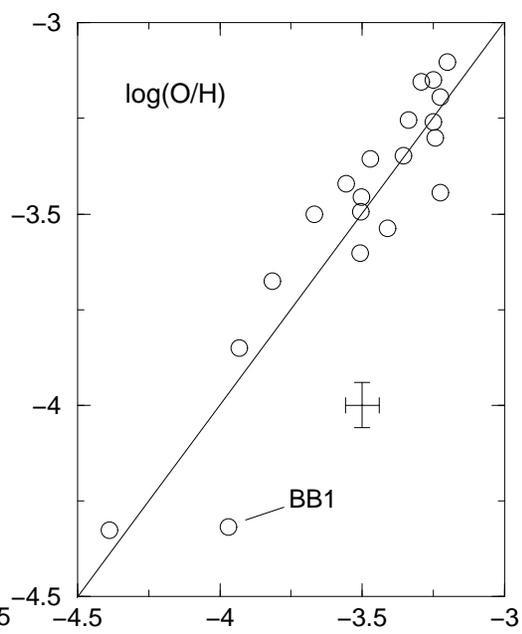
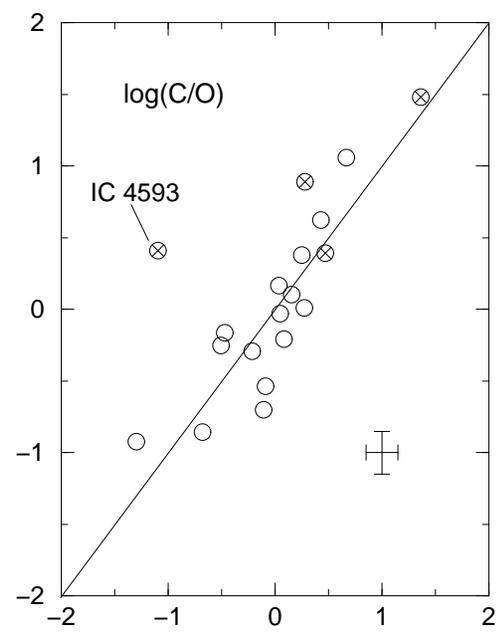
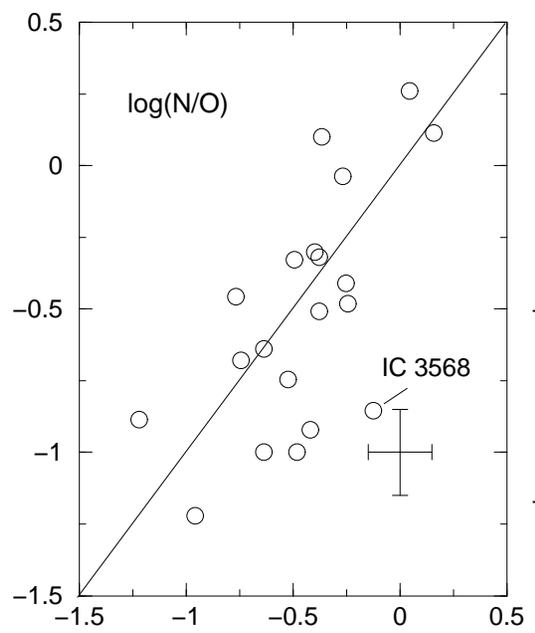
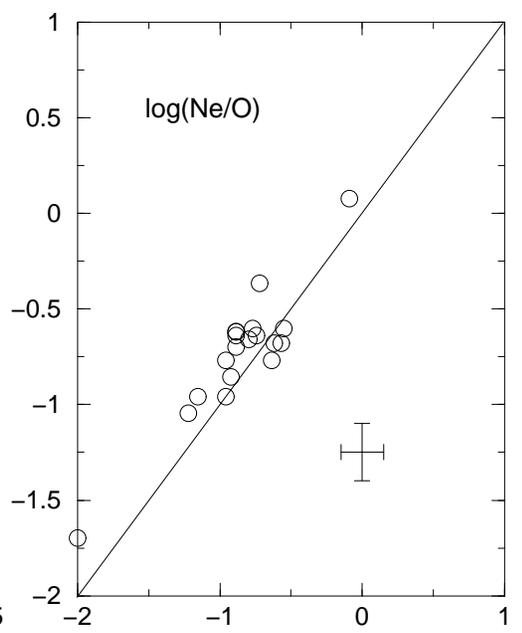
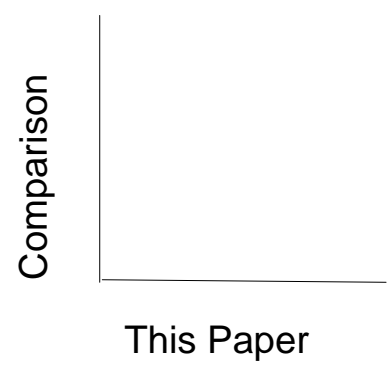
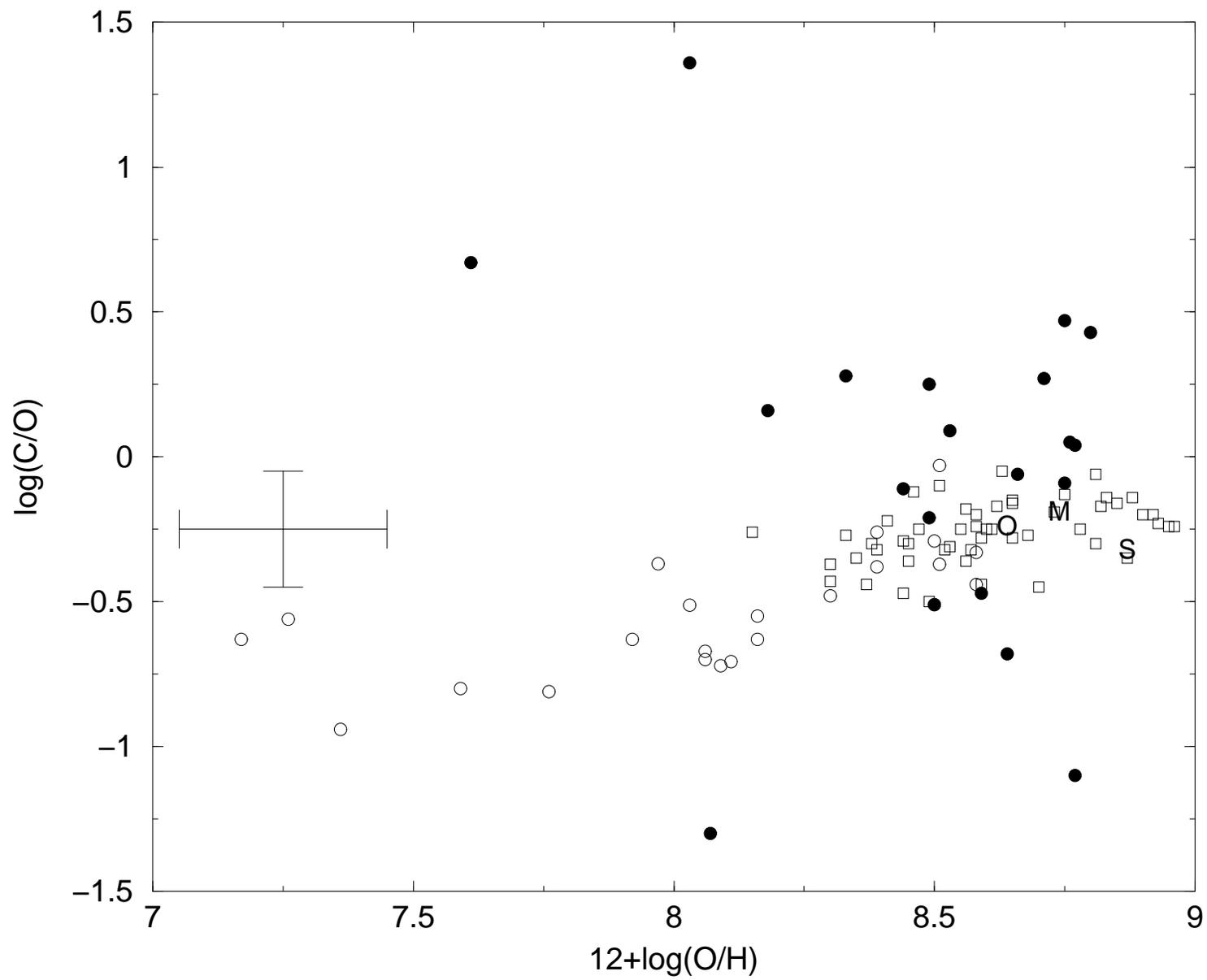
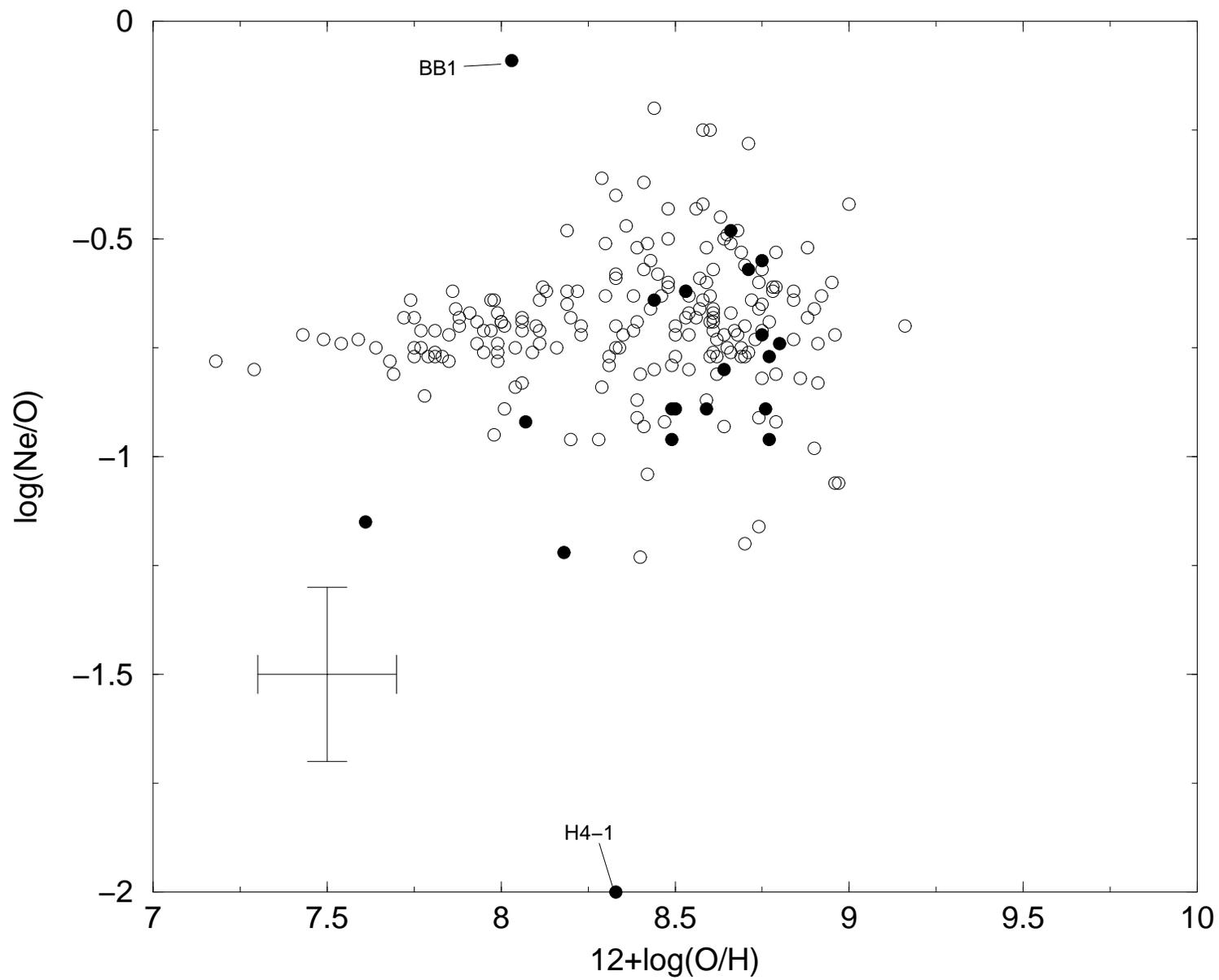
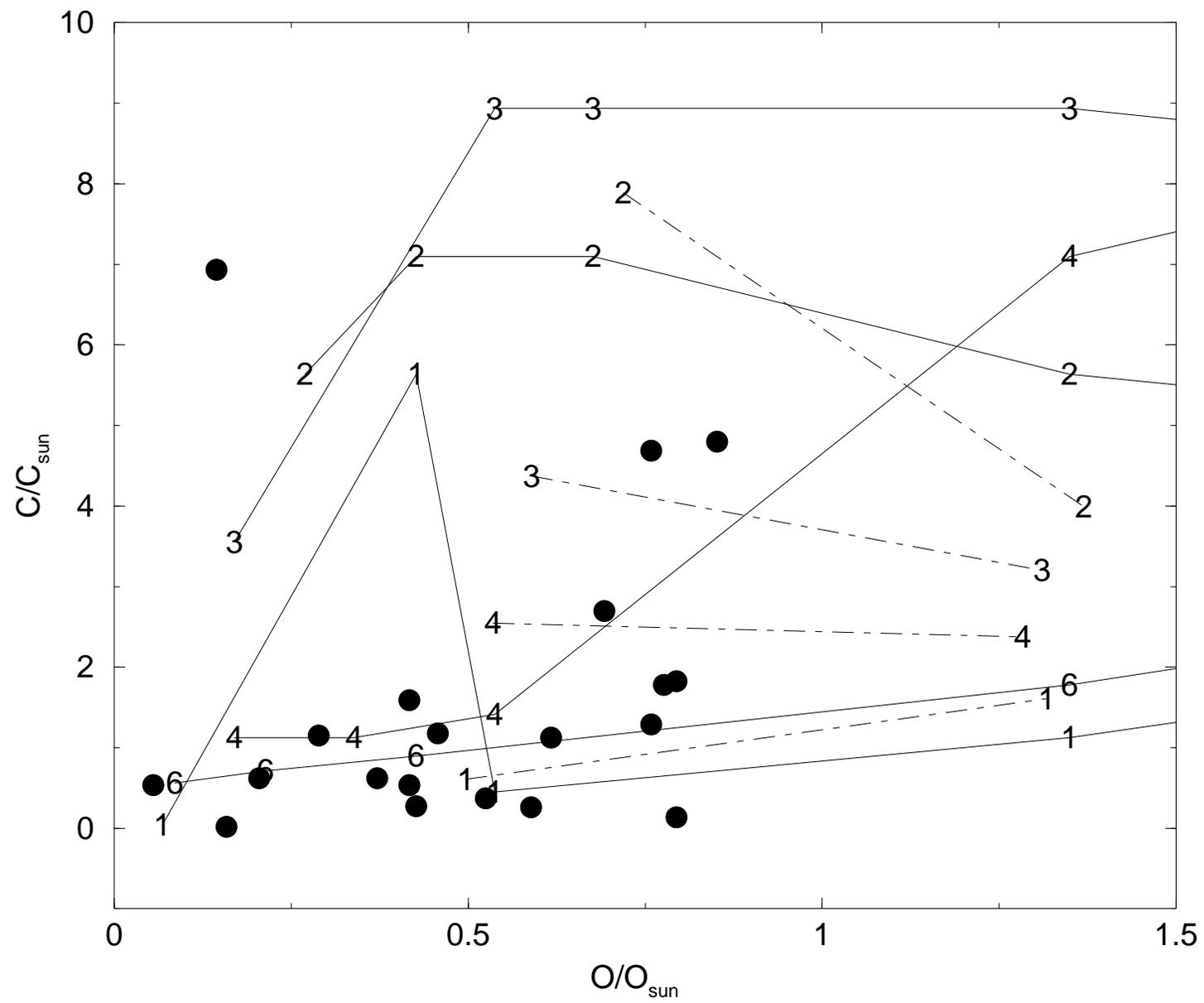
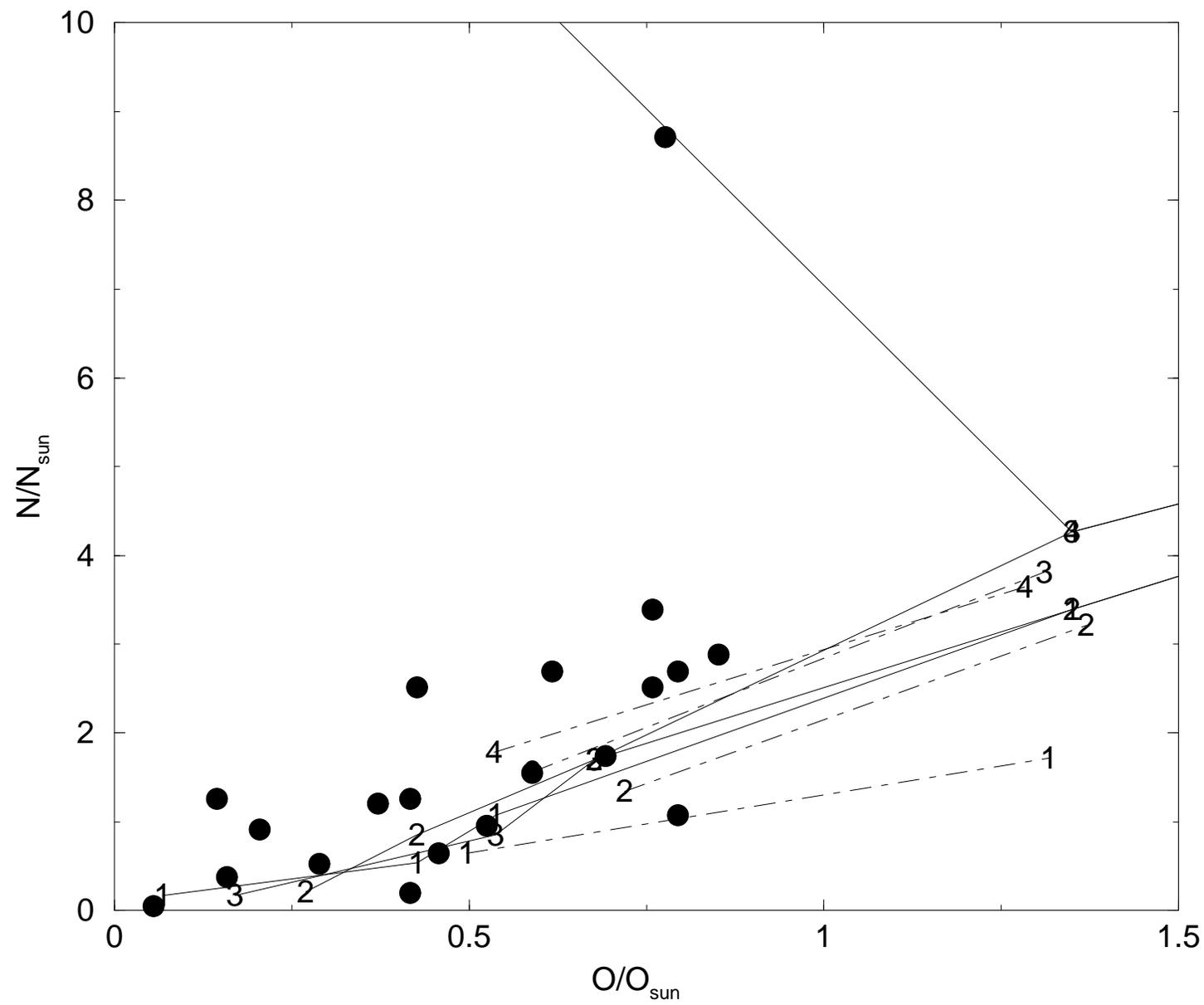
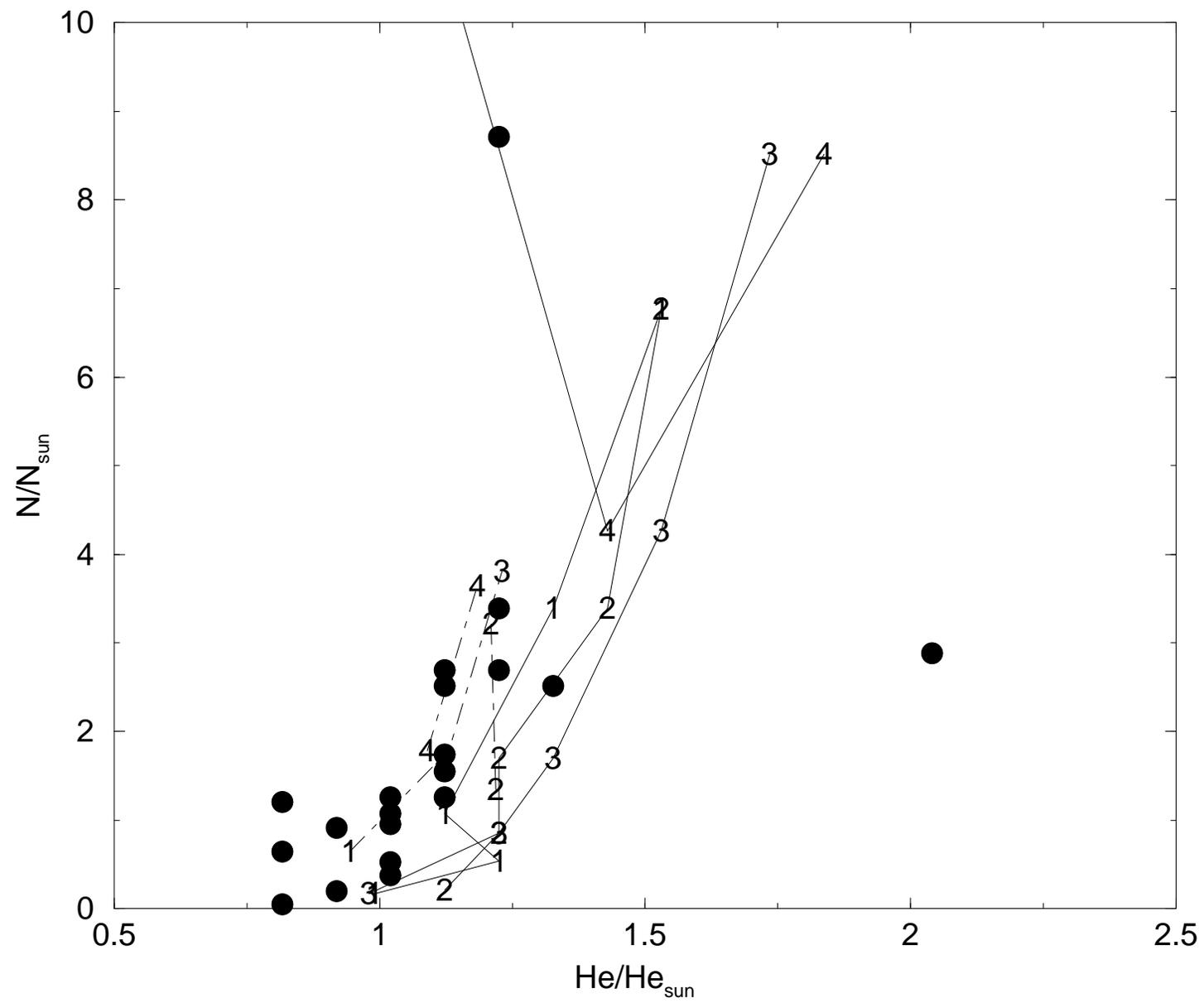
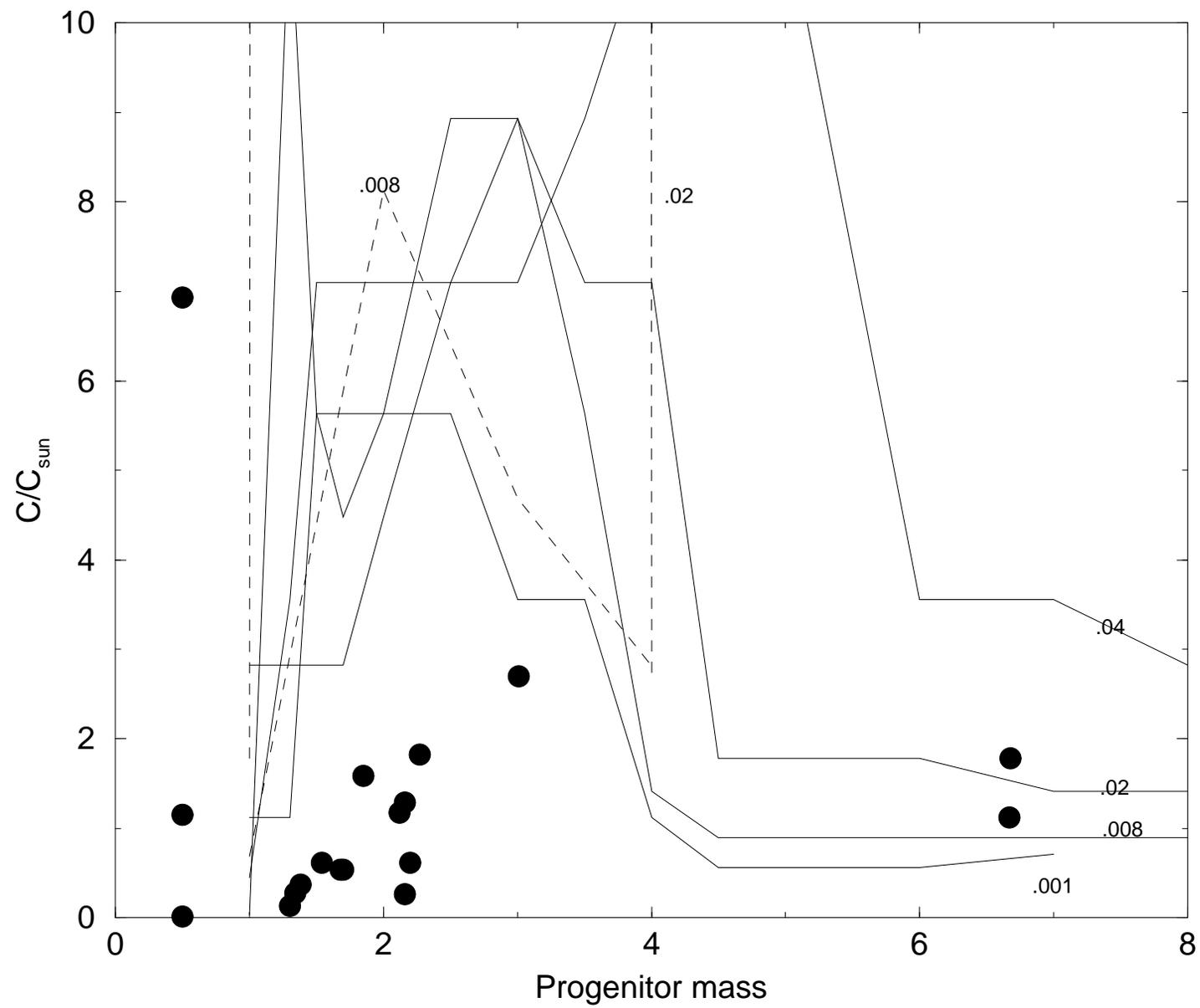
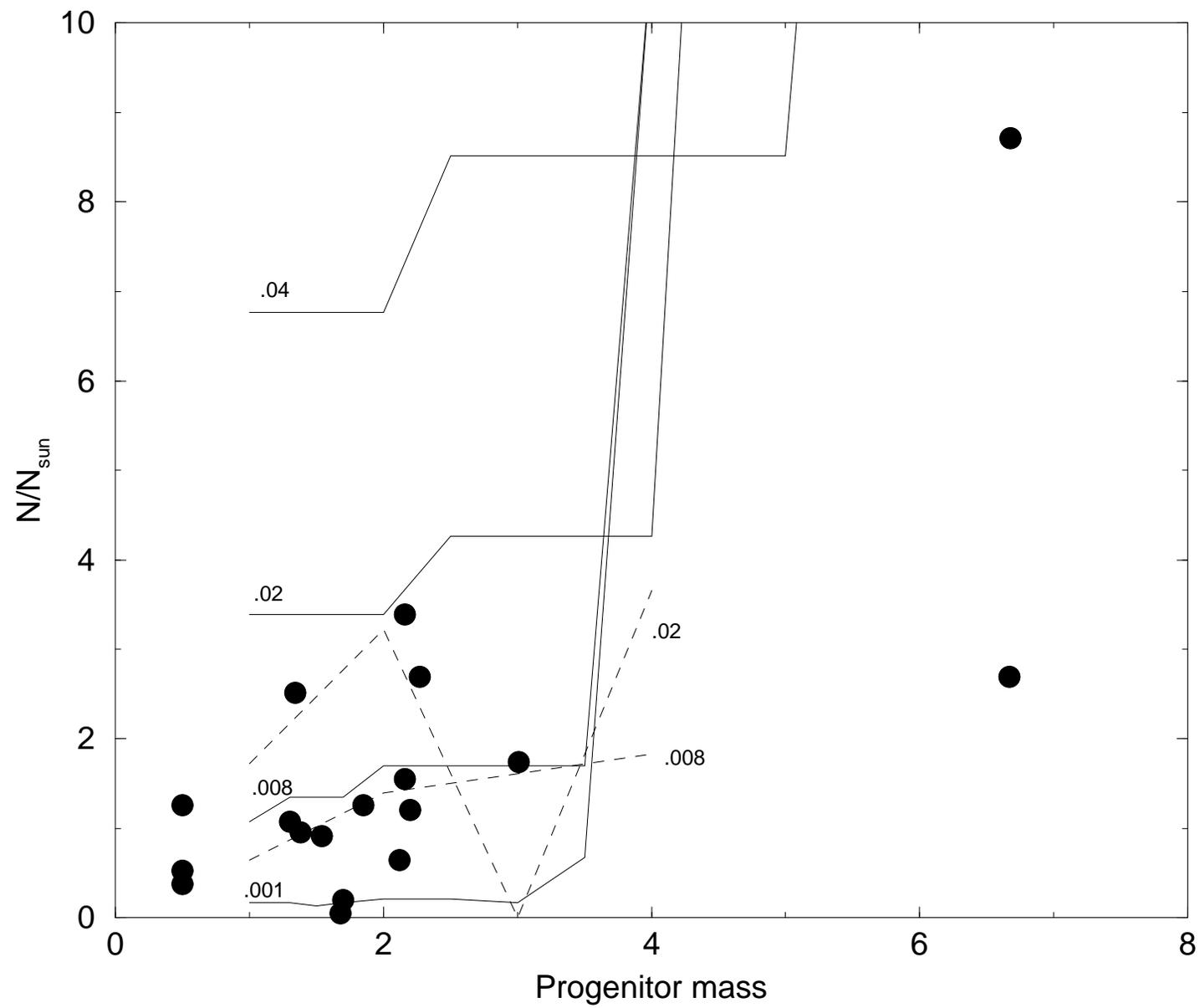
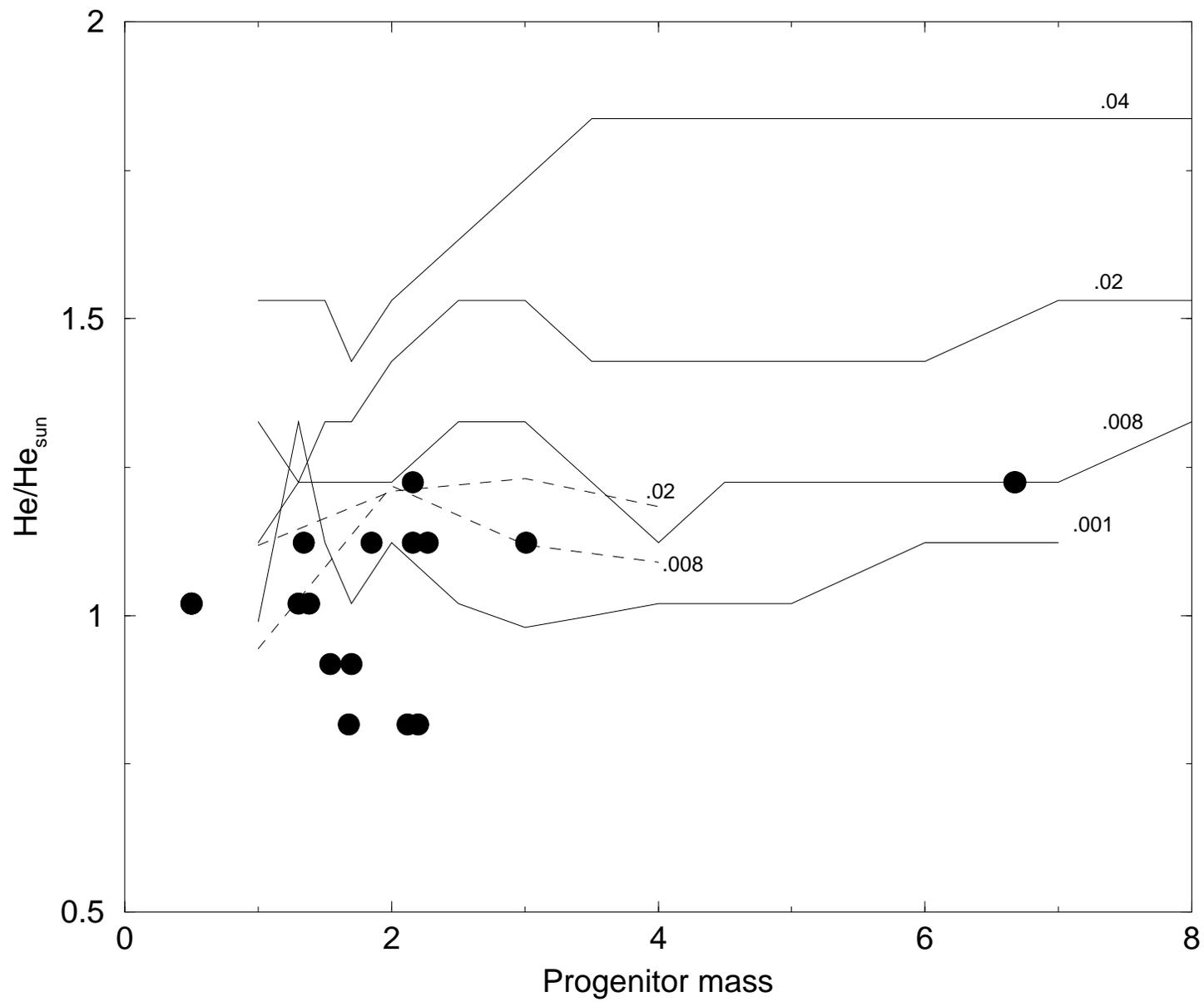
\end{document}